\documentclass{aa}

\usepackage{graphicx} 
\numberwithin{equation}{section}
\usepackage[]{hyperref}
\usepackage[]{natbib}
\usepackage{subcaption}
\usepackage{tikz}
\usepackage{pgfplots}\pgfplotsset{compat=newest}
\usepgfplotslibrary{groupplots}
\usetikzlibrary{matrix,positioning,calc,patterns,angles,quotes,arrows,external}

\newcommand{\Ha}{\mathcal H} 
\newcommand{\GravC}{\mathcal G} 

\newcommand{\D}[1]{\ensuremath{\operatorname{d}\!{#1}}} 

\newcommand{\E}{E} 
\newcommand{\M}{\ell} 
\newcommand{\N}{n} 
\newcommand{\VARPI}{\varpi} 
\newcommand{\LAMBDA}{\lambda} 

\newcommand{\LAMBDONA}{\Lambda}
\renewcommand{\P}{P}

\newcommand{\p}{p}

\newcommand{\SIGMA}{\sigma} 
\renewcommand{\SS}{S} 
\newcommand{\NU}{\nu} 
\newcommand{\NN}{N} 
\newcommand{\SIGMAONA}{\Sigma}

\newcommand{\Z}{\mathbb Z} 

\begin{document}

\title{Extreme Secular Excitation of Eccentricity Inside Mean Motion Resonance}

\subtitle{Driving Small Bodies into Star-Grazing Orbits by Planetary Perturbations}

\author{Gabriele Pichierri\inst{1},
        Alessandro Morbidelli\inst{1} \and
		Dong Lai\inst{2}
          }

\institute{	Laboratoire Lagrange, Universit\'e C\^ote d'Azur, Observatoire de la C\^ote d'Azur, CNRS, CS 34229, 06304  Nice, France
			\\ \email{gabriele.pichierri@oca.eu} \and Cornell Center for Astrophysics and Planetary Science, Department of
			Astronomy, Cornell University, Ithaca, NY 14853, USA
             }

\date{Received xxxx; accepted xxxx}

\abstract
   {It is well known that asteroids and comets fall into the Sun. Metal pollution of white dwarfs and transient spectroscopic
	signatures of young stars like $\beta$-Pic provide growing evidence that extra solar planetesimals can attain extreme orbital
	eccentricities and fall onto their parent stars.}
   {We aim to develop a general, practically implementable, semi-analytical theory of secular eccentricity excitation of small
	bodies (planetesimals) in mean motion resonances with an eccentric planet valid for arbitrary values of the eccentricities and
	including the short-range force due to General Relativity.}
   {Our  semi-analytic model for the restricted planar three-body problem does not make use of any series expansion and therefore
	is valid for any values of eccentricities and semi-major axes ratios. The model is based on the application of the adiabatic
	principle, which is valid when the precession period of the longitude of pericenter of the planetesimal is much longer than the
	libration period in the mean motion resonance. This holds down to vanishingly small eccentricities in resonances of order
	larger than 1. We provide a Mathematica notebook with the implementation of the model allowing direct use to the interested
	reader.}
   {We confirm that the 4:1 mean motion resonance with a moderately eccentric ($e'\lesssim 0.1$) planet is the most powerful one to
	lift the eccentricity of planetesimals from nearly circular orbits to star-grazing ones. However, if the planet is too
	eccentric, we find that this resonances becomes unable to pump the planetesimal's eccentricity to very high value. The
	inclusion of the General Relativity effect imposes a condition on the mass of the planet to drive the planetsimals into
	star-grazing orbits. For a planetesimal at $\sim 1$ AU around a solar-mass star (or white dwarf), we find a threshold planetary
	mass of about 17 Earth masses. We finally derive an analytical formula for this critical mass.}
   {Planetesimals can easily fall onto the central star even in the presence of a single moderately eccentric planet, but only from
	the vicinity of the 4:1 mean motion resonance. For sufficiently high planetary masses the General Relativity effect does not
	prevent the achievement of star-grazing orbits.
}


\titlerunning{Extreme secular excitation in mean motion resonances}
\authorrunning{G.\ Pichierri, A.\ Morbidelli \& D.\ Lai}

\maketitle

\section{Introduction}
In the last 30 years it has become clear that planetary perturbations can force asteroids into such highly eccentric orbits that 
they collide with the Sun. 
There is also growling evidence that planetesimals may fall onto their parent stars or suffer tidal disruption.

In the solar system, Sun-grazing long-period comets (e.g. the famous Kreutz group; \cite{1967AJ.....72.1170M}), 
have been known for  a long time, but these objects are expected to come from the Oort cloud on orbits with already 
very large eccentricities, so that planetary perturbations only play a minor role in driving their final Sun-grazing eccentricities.
But in 1994, \cite{1994Natur.371..315F}, following the evolution of the known Near-Earth objects with numerical simulations, 
discovered that asteroids frequently collide with the Sun. 
The original source of Near-Earth asteroids is the asteroid belt, 
so in this case planetary perturbations must play the major role in removing the object's initial angular momentum. 
Mean motion resonances with Jupiter and a secular resonance with Saturn were identified to be the main mechanisms capable 
of pushing the asteroid's eccentricity to large values, far more effective than planetary close encounters. 
\cite{1997Sci...277..197G}, again with numerical simulations, showed that more than 70\% of the objects initially in the 
$\nu_6$ secular resonance with Saturn or the 3:1 mean motion resonance with Jupiter eventually collide with the Sun. 

The collision of small bodies with the central star is not an oddity of our Solar System. \cite{1987A&A...185..267F} and 
\cite{1987A&A...173..289L} proposed that the red-shifted Ca II and NA I absorption lines observed in $\beta$ Pictoris 
were due to infalling evaporating bodies (see also \cite{1989A&A...223..304B,1990A&A...236..202B,1991A&A...241..488B}).
The frequency of such events, with a characteristic timescale of a few years, suggested that the infalling bodies 
were on short-period orbits, similar to asteroids or short-period comets in the solar system. 
In recent years, many more young star systems have been observed to possess Doppler-shifted, transient absorption line features 
similar to $\beta$ Pic, suggesting that infalling small bodies may be a common phenomenon
(e.g., \cite{1996A&A...309..155S,2013PASP..125..759W,2016MNRAS.461.3910G}).

Additional evidence for planetesimals falling onto the central star comes from the atmospheric pollution in heavy elements 
of white dwarfs (see  Farihi, 2016, for a review). Spectroscopic study of a large sample of cool, 
hydrogen-rich white dwarfs has established a minimum frequency of 30\% for the pollution phenomenon in these objects
(\cite{2003ApJ...596..477Z,2014A&A...566A..34K}). 
In cold white dwarfs, heavy elements should rapidly sink (\cite{1979ApJ...231..826F,1979ApJ...230..563V}) 
leaving behind only hydrogen or helium. Thus external sources must be responsible for any photospheric metals. 
The most commonly accepted explanation is that these metals originate from tidally disrupted planetesimals 
(\cite{2002ApJ...572..556D,2003ApJ...584L..91J}). 
In essence, planetesimals perturbed  into highly eccentric orbits pass within the stellar Roche limit 
(which is of the order of the solar radius $R_\odot$) and are torn apart 
by gravitational tides; subsequent collisions reduce the fragments to dust; the latter produce an infrared excess 
and slowly rain down onto the stellar surface, which generates the observed atmospheric pollution. 
Obviously, for this model to work, planetesimals have to be ``pushed'' by some planetary perturbations 
to achieve orbits that are eccentric enough to pass within $\sim R_\odot$ from the star. 
Given the ubiquity of the white dwarf pollution phenomenon, a robust mechanism of extreme eccentricity
excitation of planetesimals is needed (e.g.\ \cite{2011MNRAS.414..930B,2012ApJ...747..148D,2017ApJ...834..116P}).

These astrophysical contexts revive the interest in mean motion resonances with eccentric planets as a generic mechanism 
for pumping the eccentricities of small bodies from $\sim 0$ to $\sim 1$, i.e. for driving planetesimals into the central star. 

Analytic celestial mechanics shows that mean motion resonances with a planet on a {\it circular} orbit 
only cause an oscillation of the small body's semi-major axis coupled with a moderate oscillation of the eccentricity 
and with the libration of the angle $k\LAMBDA-k'\LAMBDA'$ (where $\LAMBDA$ and $\LAMBDA'$ are the mean longitudes 
of the small body and of the planet respectively and the integer coefficients $k$ and $k'$ define the $k':k$ resonance; 
\cite{1983CeMec..30..197H,1984CeMec..32..109L}). 
However, if the perturbing planet has some finite eccentricity, inside a mean motion resonance there can be a 
dramatic secular evolution, with the eccentricity of the small body undergoing large excursions correlated with 
the precession of the longitude of perihelion 
(\cite{1985Icar...63..272W,1983Icar...56...51W,1990CeMDA..47...99H}). 

These pioneer works used a series expansion of the Hamiltonian in power laws of the eccentricities 
of the perturbed body ($e$) and of the planet ($e'$), and focused specifically on the case of the 3:1 resonance with Jupiter. 
A few years later, \cite{1993CeMDA..57...99M,1995Icar..114...33M} developed a semi-analytic study of the dynamics 
in mean motion resonances using a first order expansion in $e'$ but no series expansions in the $e$. 
This way, they could follow the evolution of the small body to arbitrary larger eccentricities. 
This approach is valid only for small values of $e'$ and for $e>e'$. 
Motivated by the \cite{1994Natur.371..315F} numerical results, Moons and Morbidelli focused on the specific case 
of the Solar System, including the effects of Saturn on the orbital evolution of Jupiter in addition to their combined 
perturbation to the asteroid. 
In this framework, they established the existence of overlapping secular resonances inside the 4:1, 3:1, 5:2 and 7:3 
mean motion resonances, which can push the eccentricity of the small body to unity. 

In a more general context, \cite{1996Icar..120..358B} investigated the secular dynamics in mean motion resonances 
with a single planet with various (albeit moderate) eccentricities. Again, they considered an expansion in $e'$ to first order, 
and no expansion in the eccentricity of the perturbed body. 
They found that, of all resonances, the 4:1 is the most powerful in pushing the eccentricity 
of the small body from $\sim 0$ to $\sim 1$, provided that $e'\gtrsim 0.05$. 
By contrast, the 3:1 resonance only generates large oscillations in the eccentricity of the small body, 
but insufficient to produce star-grazing orbits, at least for planet's eccentricities up to 0.1. 
Because of the linear expansion in $e'$, \cite{1996Icar..120..358B} could not determine the threshold planetary eccentricity 
in order to generate the star-grazing phenomenon for small bodies initially on quasi-circular orbits in the 3:1 resonance.

In this paper we revisit the problem of the eccentricity evolution of small bodies inside mean motion resonances with 
an eccentric planet using a semi-analytic approach. 
In order to go beyond the previous works, we do not expand the Hamiltonian in the eccentricity 
of either the small body or the perturber. 
In this way, our study is valid for all eccentricities and also in the $e<e'$ regime. 
Our work is not the first to avoid expansions in $e'$ (e.g.\ \cite{2006MNRAS.365.1160B,2006CeMDA..94..411M}), 
but it is the first to do so for the problem of secular 
evolution of a small body in mean motion resonance with a planetary perturber.
We use the adiabatic principle (already invoked in \cite{1985Icar...63..272W}) to disentangle the motion related to the libration 
of $k\LAMBDA-k'\LAMBDA'$ from the secular motion relating eccentricity and longitude of perihelion. 
To remain relatively simple, our analysis is performed in the limit of small amplitude of libration in the mean motion resonance. 

The paper is structured as follows. 
In Section 2, we develop the analytic formalism for the study of the secular dynamics at the core of mean motion resonances, 
without series expansions. This results in a two degree-of-freedom averaged Hamiltonian \eqref{eq:AveragedHamiltonian}.  
In Section 3 we lay out the method for studying the dynamics given by the averaged Hamiltonian, 
using the theory of adiabatic invariance; we also discuss the limit of validity of this method.
In Section 4 we also include a post-Newtonian term, describing the fast precession of the longitude of perihelion 
at large eccentricity due to General Relativity. 
Our results are presented in Section 5. We first neglect the General Relativity effect, in which case the secular evolution is 
independent of the planet's mass, only the timescale of the secular evolution depends on it. 
We focus in particular on the 4:1, 3:1 and 2:1 resonances and, for each of these resonances, 
we evaluate what planetary eccentricities are needed for lifting bodies from initially 
quasi-circular orbits to star-grazing ones, if it is ever possible. 
When this is the case, we then introduce the post-Newtonian correction, 
which makes the secular dynamics at high eccentricity dependent on the planetary mass. Thus we evaluate, 
for the resonances and the planetary eccentricities previously considered, what is the minimal planetary mass required 
to achieve the star-grazing phenomenon. 
In addition, we provide a Mathematica notebook implementing our model, available at \url{www.oca.eu/morby/SecResInMMR.nb}, 
so that the reader can compute the secular dynamics in the desired resonances with the desired planets. 
The conclusions of this work are summarized in Section 6.
 
\section{The planetary Hamiltonian}

We start with the Hamiltonian for the restricted planar three-body problem. 
By denoting with $\mathbf{x} = (x,y)$ and $\mathbf{v} = (v_x, v_y)$ the Cartesian coordinates and momenta of 
the perturbed test particle (``small body''), and using a prime for the perturber (``planet''), the Hamiltonian reads:
\begin{equation}\label{eq:HR3BP}
\begin{split}
\Ha &= \Ha_{kepl} + \Ha_{pert}\\
	&= \frac{\|\mathbf v\|^2}{2} - \frac{\GravC M_*}{\|\mathbf x\|} - 
		\GravC m' \left(\frac{1}{\Delta} - \frac{\mathbf x \cdotp \mathbf x'}{\|\mathbf x'\|^3} \right),
\end{split} 
\end{equation}
where $M_*$ is the mass of the star, $m'$ is the mass of the perturber, and
$\Delta = \| \mathbf x - \mathbf x' \|$ is the distance between the test particle and the perturber (\cite{2000ssd..book.....M}). 
The perturber is assumed to follow a given Keplerian orbit, so $\mathbf x'$ is a function of time.
In terms of orbital elements, the Cartesian coordinates are given by
\begin{equation}\label{eq:CartesianComponents}
\begin{split}
x &= a (\cos{\E} - e) \cos{\VARPI} - 
   a \sqrt{1 - e^2} \sin{\E} \sin{\VARPI}, \\
y &= a (\cos{\E} - e) \sin{\VARPI} + 
    a \sqrt{1 - e^2} \sin{\E} \cos{\VARPI},
\end{split}
\end{equation} 
where $a$ is the semi-major axis, $e$ is the eccentricity and $\E$ is the eccentric anomaly of the
perturbed, and similar (primed) equations for the perturber (\cite{2000ssd..book.....M}).

We introduce the canonical modified Delaunay action-angle variables $(\LAMBDONA, \P, \LAMBDA, \p)$, given by
\begin{alignat}{2}\label{eq:ModifDelaunayVariables}
\LAMBDONA 	&=\sqrt{\GravC M_* a},						&&\quad \LAMBDA = \M + \VARPI, \nonumber\\
\P			&=\sqrt{\GravC M_* a}(1-\sqrt{1-e^2}),	 	&&\quad \p = -\VARPI,
\end{alignat}
where $\LAMBDA$ is the mean longitude, $\M = \E - e\sin\E$ is the mean anomaly and $\VARPI$ is the longitude of 
pericenter of the test mass.
In order to make the system autonomous, we extend the phase space by introducing for the perturber
\begin{alignat}{2}
\LAMBDONA',	&&\quad\quad\quad \LAMBDA' = \M' + \VARPI' = \N' (t-t_0),
\end{alignat}
where $\N' = \sqrt{\GravC M_*/a'^3}$ is the mean motion of the perturber. 
We assume that the perturber does not precess, so without loss of generality we set $\VARPI' = 0$.
Now the autonomous Hamiltonian of the system reads
\begin{equation}
\begin{split}
\Ha(\LAMBDONA,\P, \LAMBDONA',\LAMBDA,\p, \LAMBDA') = &-\frac{\GravC ^2 M_*^2}{2 \LAMBDONA^2} + \N' \LAMBDONA' + \\ 
									&+\Ha_{pert}(\LAMBDONA,\P, \LAMBDONA',\LAMBDA,\p, \LAMBDA'; e', \VARPI' = 0),
\end{split}
\end{equation}
where the perturbation part $\Ha_{pert}$ is to be written in terms of the newly defined variables. 
We have stressed that it depends parametrically on the arbitrary values of $e'$ and $\VARPI'=0$.

We now consider the test particle to be (close to an) inner mean motion resonance with the outer perturber. In other
words, we assume $k \N - k' \N' \sim 0$, where $\N=\sqrt{\GravC M_*/a^3}$ is the mean motion of the test particle,
for some positive integers $k$, $k'$, such that $k'>k$.
In order to study the resonant dynamics, one may introduce a set of canonical resonant action-angle variables:
\begin{alignat}{2}\label{eq:ResonantActionAngleVariables}
\SS	&=P,										&&\quad \SIGMA = \frac{k'\LAMBDA' - k\LAMBDA+(k'-k)\p}{(k'-k)}, \nonumber\\
\NN	&=\frac{k'-k}{k}\LAMBDONA + P,				&&\quad \NU = \frac{-k'\LAMBDA' + k\LAMBDA}{(k'-k)} = - \SIGMA + \p,\\
\tilde\LAMBDONA' &= \LAMBDONA'+ \frac{k'}{k}\LAMBDONA,	&&\quad \tilde\LAMBDA'=\LAMBDA'. \nonumber
\end{alignat}
The historical reason for adopting these variables is that for $e'=0$ there is no harmonic term in $\NU$ in the 
Hamiltonian and thus $\NN$ is a constant of motion.
The reason why the critical resonant angle $\SIGMA$ is not simply defined as $k'\LAMBDA' - k\LAMBDA+(k'-k)\p$
is explained in \cite{1984CeMec..32..109L}. 
That is, because of the d'Alembert rules, the coefficient of the terms $\cos[l(k'\LAMBDA' - k\LAMBDA+(k'-k)\p)]$ 
in the Fourier expansion of the perturbing Hamiltonian is proportional to $e^{l|k'-k|}$ for small values of $e$. 
Thus for small eccentricities the Hamiltonian is a polynomial expression in $e\cos\SIGMA$ and $e\sin\SIGMA$, 
and the apparent singularity at $e=0$ can be removed.

Using the variables \eqref{eq:ResonantActionAngleVariables}, the Keplerian part of the Hamiltonian takes the form
\begin{equation}\label{eq:KeplerianHamiltonianInResonantVariables}
\begin{split}
\Ha_{kepl}(\SS, \NN, \tilde\LAMBDONA') = &-\GravC^2 M_*^2 \frac{ (k'-k)^2}{2 k^2 (\NN-\SS)^2} +\\ 
				&+\N'\left[ \tilde\LAMBDONA' - \frac{k'}{(k'-k)}(\NN-\SS)\right].
\end{split}
\end{equation}

At this point, we average the Hamiltonian over the fast angles. 
From a computational point of view, a remark is in order. 
The Cartesian components given in \eqref{eq:CartesianComponents}
are expressed in terms of the eccentric anomalies $\E$, $\E'$. 
Thus it would be necessary to invert the Kepler's equation
$\LAMBDA - \VARPI = \M = \E - e\sin\E$ to obtain $\E = \E(\LAMBDA)$, and similarly for $\E'=\E'(\LAMBDA')$. 
If $e'$ is not too large, the latter inversion is not problematic. 
However the eccentricity $e$ of the test particle can reach high values,
so that solving the Kepler equation for the test particle becomes numerically cumbersome. 
Therefore, it is advisable to retain the dependence on the 
eccentric anomaly $\E$, use the differential relationship $\D\LAMBDA = (1-e\cos\E)\D\E$, and integrate over $\E$ instead. 
This is more convenient because the dependence of $\LAMBDA$ on $\E$ is given through Kepler's equation by an explicit formula.
Also note that $\LAMBDA$ is related to $\LAMBDA'$ through $\SIGMA$ by $\LAMBDA'=[(k\LAMBDA-(k'-k)(\p-\SIGMA)]/k'$. 
In summary, using the resonant relationship and Kepler's equation one obtains
$\E'$ from $\LAMBDA'$, $\LAMBDA'$ from $\LAMBDA$ and $\LAMBDA$ from $\E$, so 
that the averaging or $\E$ eliminates the short periodic dependence of the
Hamiltonian. 
By doing so, the canonical angle $\LAMBDA'$ vanishes from the averaged Hamiltonian, 
and $\tilde\LAMBDONA'$ becomes a constant of motion, so that the term $\N' \tilde\LAMBDONA'$ 
can be dropped from \eqref{eq:KeplerianHamiltonianInResonantVariables}. 

Proceeding this way, we have that
\begin{equation}\label{eq:AveragedPerturbationHamiltonian}
\bar\Ha_{pert}(\SS,\NN,\SIGMA,\NU):=\frac{1}{2\pi k'} \int_{0}^{2\pi k'} \Ha_{pert} \cdotp (1-e\cos\E)\D\E;
\end{equation}
note that we integrate over $\E$ from $0$ to $2\pi k'$ instead of just $2\pi$ because only after $k'$ revolutions 
of the test particle around the star (which correspond to $k$ revolutions of the outer perturber) 
does the system attain the initial configuration, thus recovering the complete periodicity of the Hamiltonian.  
The integral \eqref{eq:AveragedPerturbationHamiltonian} can be solved numerically. 
In our code we use
a Mathematica function with an imposed relative accuracy of $10^{-10}$.
For the Keplerian part we just write
\begin{equation}\label{eq:AveragedKeplerianHamiltonian}
\bar\Ha_{kepl}(\SS,\NN):=-\GravC^2 M_*^2 \frac{(k'-k)^2}{2 k^2 (\NN-\SS)^2} - 
				\N' \frac{k'}{(k'-k)}(\NN-\SS).
\end{equation}
The averaged Hamiltonian then becomes
\begin{equation}\label{eq:AveragedHamiltonian}
\bar\Ha (\SS,\NN,\SIGMA,\NU) := \bar\Ha_{kepl} + \bar\Ha_{pert}.
\end{equation}
This two degree-of-freedom system is not integrable in general, unless further approximation is made.

\section{Studying the averaged Hamiltonian}\label{sec:AdiabaticMethod}

We now intend to study quantitatively the dynamics given by the Hamiltonian \eqref{eq:AveragedHamiltonian}. 
This can be seen as an integrable system (i.e.\ the Keplerian part), to which a small perturbation is added, 
of order $\mu = m'/M_* \ll 1$.

We begin by noticing that $\bar\Ha_{kepl}$ depends on $\NN-\SS$ only, so it is convenient to introduce the canonical 
variables
\begin{alignat}{2}
&\SIGMAONA = \SS-\NN,	&&\quad \SIGMA, \nonumber\\
&\NN,					&&\quad \p = \SIGMA+\NU,
\end{alignat}
making $\bar\Ha_{kepl}$ a function of $\SIGMAONA$ only.
The location of exact resonance is given by the value $\SIGMAONA=\SIGMAONA_{res}$ such that
\begin{equation}\label{eq:LocationOfExactResonance}
\frac{\partial \bar\Ha_{kepl}}{\partial \SIGMAONA} (\SIGMAONA_{res})=0, \quad \text{i.e.} ~~ 
	\SIGMAONA_{res} = - (\GravC M_*)^{2/3} \frac{(k'-k)}{(k^2 k' n')^{1/3}}, 
\end{equation}
which is nothing but $\N=\N_{res}=(k'/k)\N'$. 
The expansion of $\bar\Ha_{kepl}$ in $\Delta\SIGMAONA=\SIGMAONA-\SIGMAONA_{res}$ starts with a quadratic term in $\Delta\SIGMAONA$.
Since the perturbation $\bar\Ha_{pert}$ is a function of $(\SIGMAONA,\NN,\SIGMA,\p)$ and is of order $\mu$,
the dynamics in the canonical pair of variables $(\SIGMAONA,\SIGMA)$ near $\SIGMAONA_{res}$ is equivalent to that of 
a pendulum with Hamiltonian of the form $(\Delta\SIGMAONA)^2 + \mu \cos\SIGMA$, so its frequency is of order $\sqrt\mu$.
On the other hand, the dynamics in the canonical pair $(\NN,\p)$ is slower, with a characteristic frequency of order $\mu$.
We can therefore apply the adiabatic principle and study the dynamics in $(\SIGMAONA,\SIGMA)$ with fixed $(\NN,\p)$,
and then the dynamics in $(\NN,\p)$ keeping constant the action integral 
\begin{equation}\label{eq:AdiabaticInvariant}
J=\oint \SIGMAONA \D\SIGMA,
\end{equation} 
which is the adiabatic invariant of the dynamics (\cite{Henrard1993}).

\begin{figure*}[!ht]
\centering
\begin{subfigure}[b]{0.45 \textwidth}
\centering
\includegraphics[scale=0.6
]{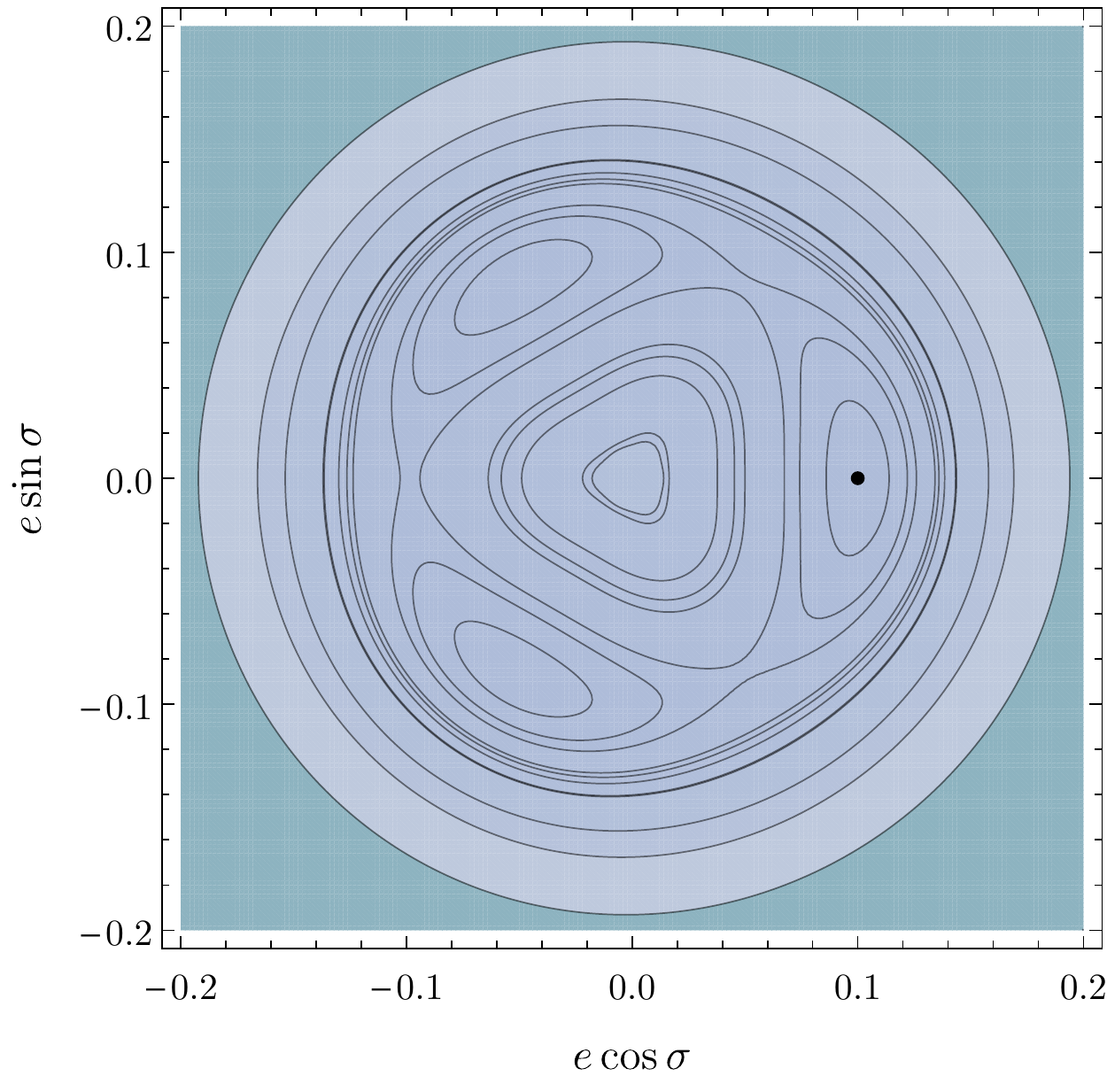}

\caption{$\NN=1.893$, $\p=0$.}
\end{subfigure}
\hfill
\begin{subfigure}[b]{0.45 \textwidth}
\centering
\includegraphics[scale=0.6
]{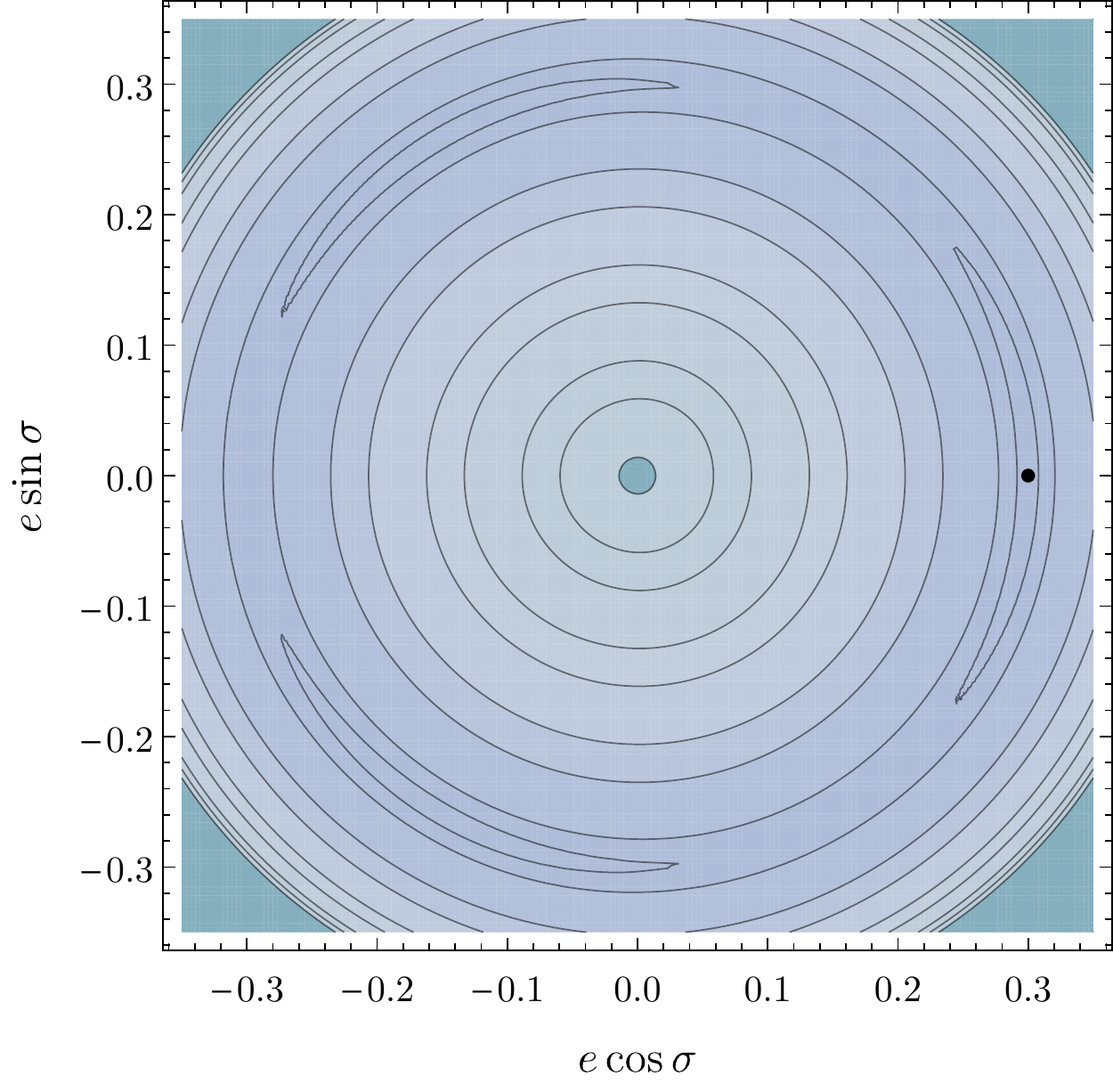}
\caption{$\NN=1.919$, $\p=0$.}
\end{subfigure}

\vskip\baselineskip

\begin{subfigure}[b]{0.45 \textwidth}
\centering
\includegraphics[scale=0.6
]{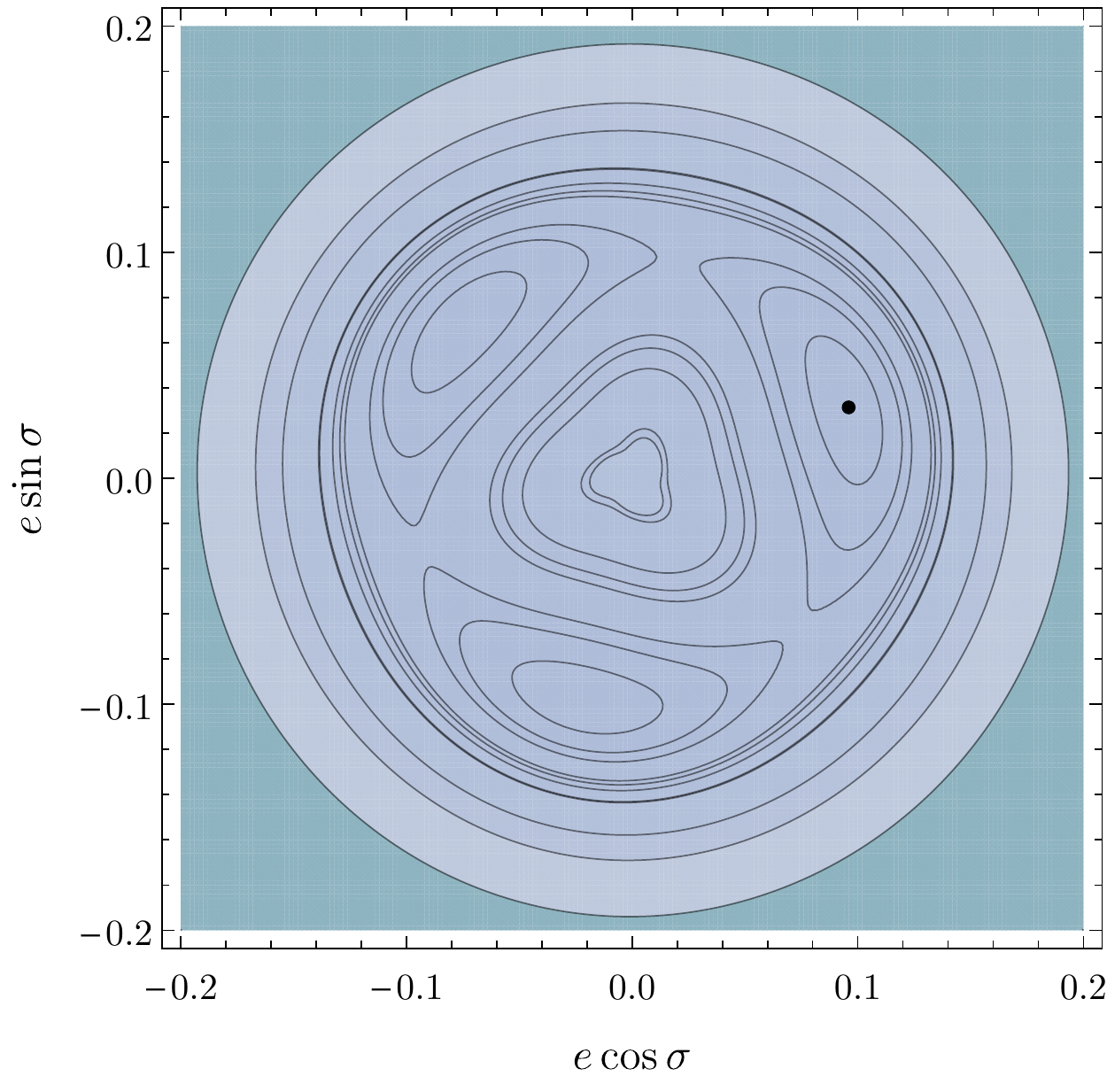}
\caption{$\NN=1.893$, $\p=\pi/8$.}
\end{subfigure}
\hfill
\begin{subfigure}[b]{0.45 \textwidth}
\centering
\includegraphics[scale=0.6
]{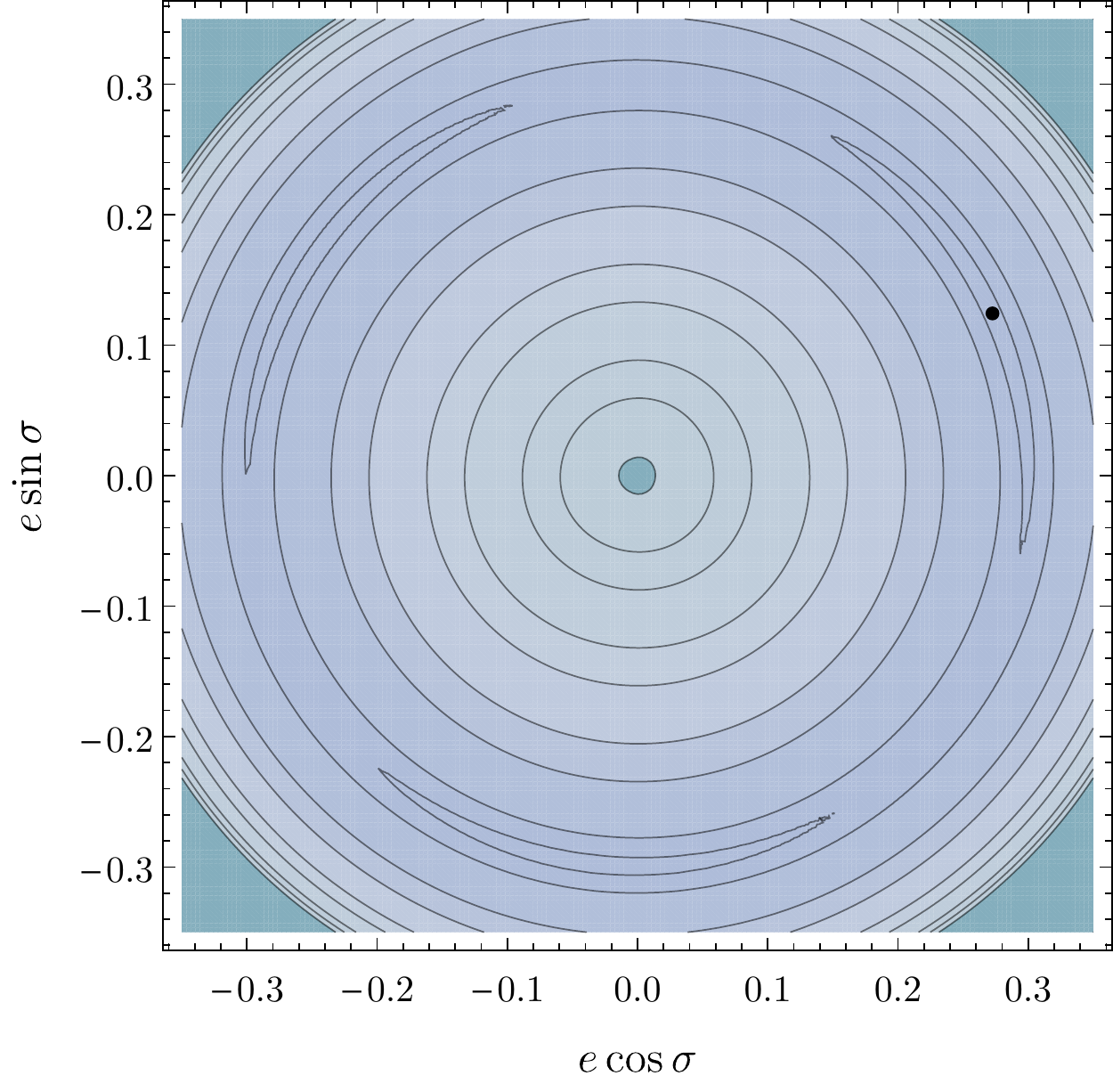}
\caption{$\NN=1.919$, $\p=\pi/8$.}
\end{subfigure}

\caption{Level plots of the Hamiltonian $\bar\Ha_{(\NN,\p)}(\SIGMAONA,\SIGMA)$ on the $(e\cos\SIGMA,e\sin\SIGMA)$
plane, for different values of $\NN$ and $\p$.
This is the case of $k'=4$, $k=1$, with $e'=0.1$, $a'=1~ \text{AU}$ and $\mu=10^{-3}$ for the perturber 
(in units where $\GravC M_*=1$). 
The black dot in each panel indicates the position of the stable equilibrium point that is found by our 
algorithm. 
Recall that the function $\bar\Ha_{(\NN,\p)}$ is periodic in $\SIGMA$ with period $2\pi/(k'-k)$, so there will always be
$k'-k$ equivalent stable equilibra one rotation away from one another.
Note how $e_{eq}$ increases with $\NN$, while $\p$ has the effect of simply rotating the diagram.}
\label{fig:Figure1} 
\end{figure*}

We now explain this procedure in more detail.
Once the values of $\NN$ and $\p$ have been fixed, $\bar\Ha$ reduces to a one degree of 
freedom Hamiltonian in $(\SIGMAONA,\SIGMA)$ and parametrized by $(\NN,\p)$, which we denote by
$\bar{\Ha}_{(\NN,\p)} (\SIGMAONA,\SIGMA)$. 
This Hamiltonian is therefore integrable, so we can study its dynamics by plotting its level curves. 
Note however that by fixing $\NN$ we can obtain $\SIGMAONA$ from $e$ and {\it vice versa},
so we can also use $(e\cos\SIGMA,e\sin\SIGMA)$ as independent variables. 
Although these variables are not canonical, they have the already mentioned advantage that for small 
$e$ the Hamiltonian is a polynomial in $(e\cos\SIGMA,e\sin\SIGMA)$, so the level curves do not have a singularity at $e=0$. 
Besides, the plot of the level curves does not require the use of canonical variables.
We show such plots in the case of the 4:1 resonance in Figure \ref{fig:Figure1}.

In principle, the dynamics can be studied for any value of $J$.
Once the cycle of $\bar{\Ha}_{(\NN,\p)}(\SIGMAONA,\SIGMA)$ corresponding to the considered value of $J$ 
through \eqref{eq:AdiabaticInvariant} is identified, the full Hamiltonian $\bar{\Ha} (\SIGMAONA,\SIGMA,\NN,\p)$ 
is averaged over such a cycle, as explained in \cite{Henrard1993}, leading to a new one-degree of freedom Hamiltonian 
$\bar{\bar{\Ha}} (\NN,\p;J)$. 
This Hamiltonian is integrable, and the resulting dynamics in $(\NN,\p)$ describes the secular evolution 
of the small body inside the mean motion resonance with the perturber.  

In this paper we simplify vastly this procedure by limiting ourselves to the case $J\to 0$, i.e.\ the limit of 
small libration amplitude in the mean motion resonance. 
In this limit, the cycle in $(\SIGMAONA,\SIGMA)$ described by $\bar{\Ha}_{(\NN,\p)}$ shrinks to the stable 
equilibrium point. 
Thus, there is no need to average the full Hamiltonian over a cycle: $\bar{\bar{\Ha}}(\NN,\p;J=0)$ is obtained 
by evaluating $\bar\Ha$ on the stable equilibrium point of $\bar{\Ha}_{(\NN,\p)}$ in the variables $e$ and $\SIGMA$. 
Note that by having fixed $\NN$, we are effectively linking the semi-major axis $a$ to the eccentricity $e$, via the relation
\begin{equation}\label{eq:aFromNNande}
a = \frac{\NN^2}{\GravC M_* \left(k'/k-\sqrt{1-e^2} \right)^2}.
\end{equation}
Therefore we recover the equilibrium values $a_{eq}$ of the semi-major axis as well. 

\begin{figure}[!ht]
\centering
\begin{subfigure}[b]{0.45 \textwidth}
\centering
\includegraphics[scale=0.45
]{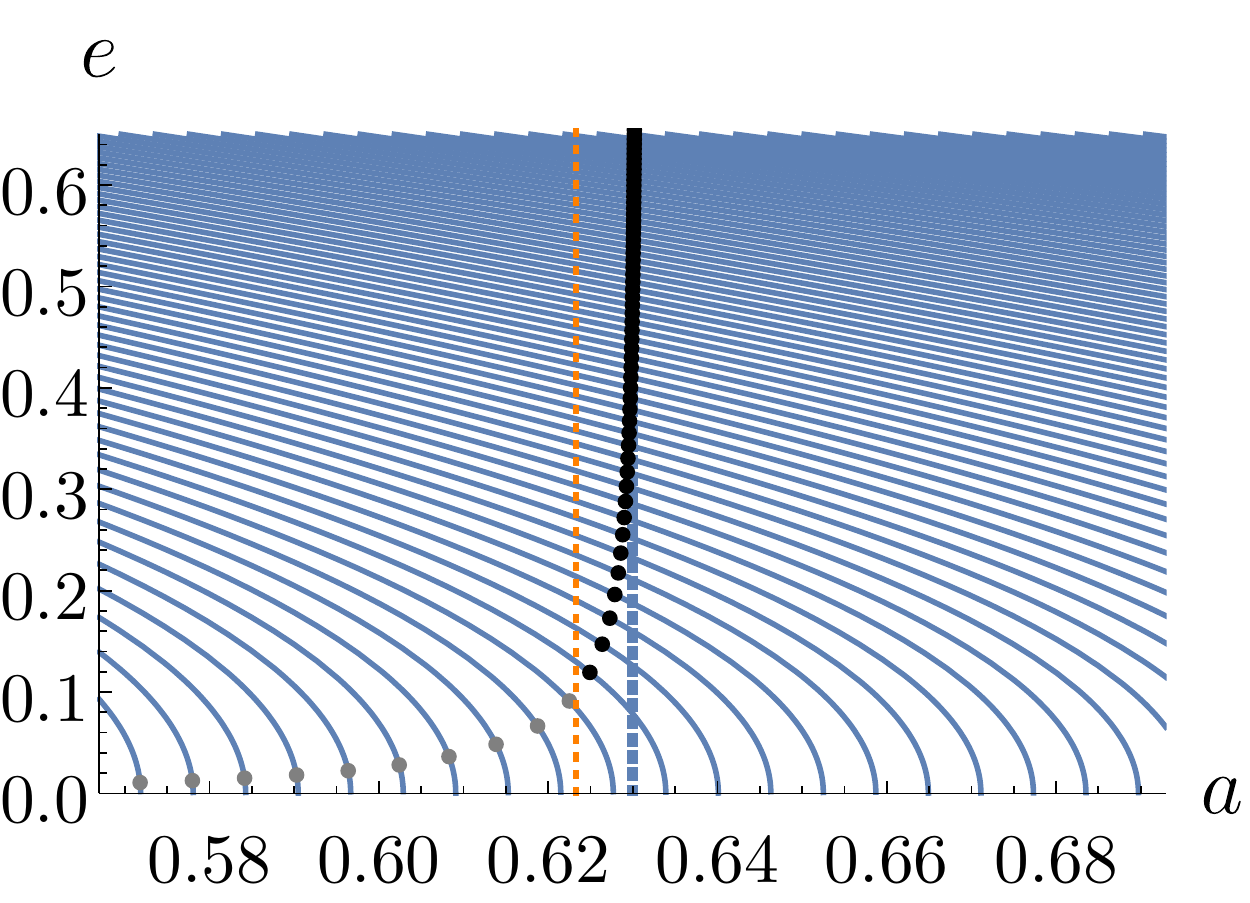}
\caption{$2:1$ resonance.}
\end{subfigure}

\begin{subfigure}[b]{0.45 \textwidth}
\centering
\includegraphics[scale=0.45
]{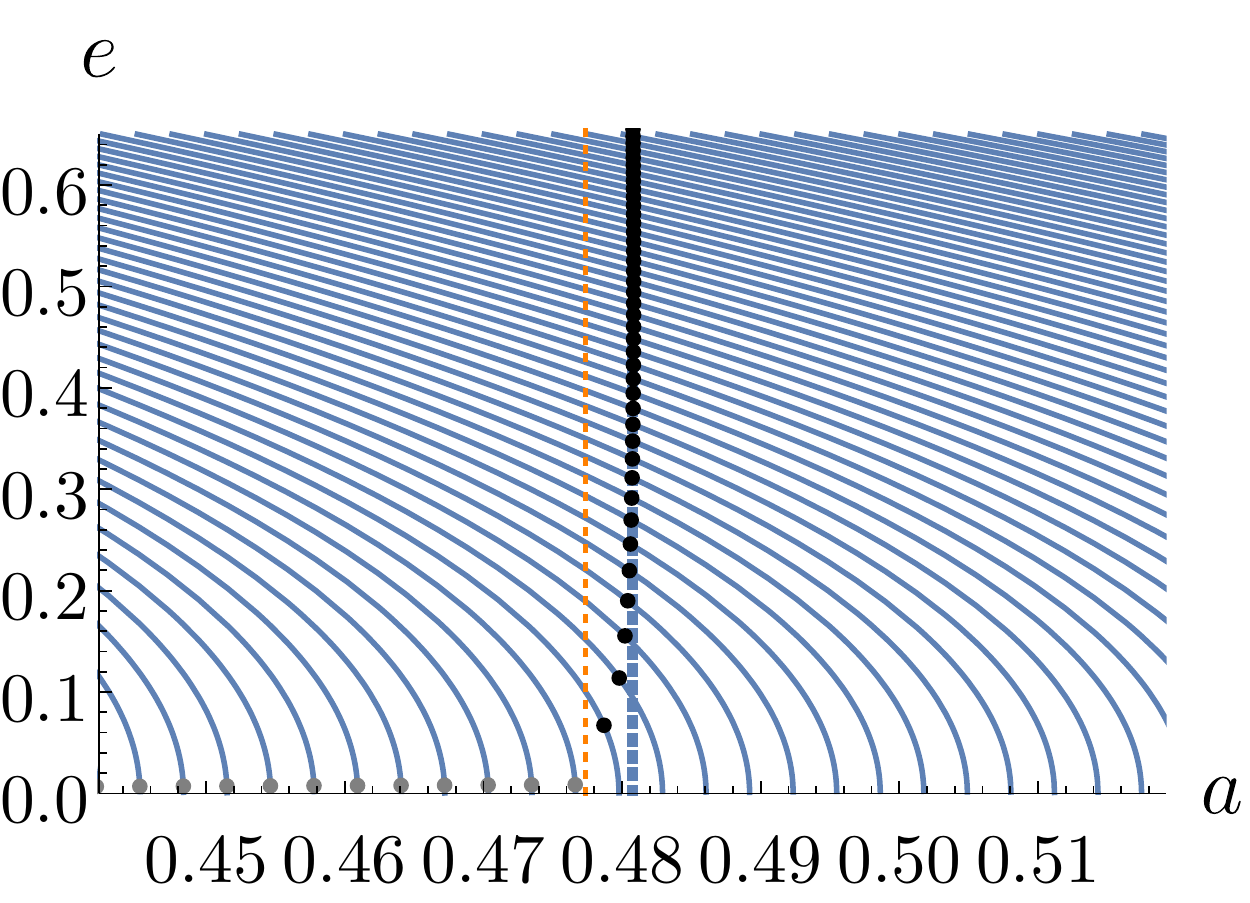}
\caption{$3:1$ resonance.}
\end{subfigure}

\begin{subfigure}[b]{0.45 \textwidth}
\centering
\includegraphics[scale=0.45
]{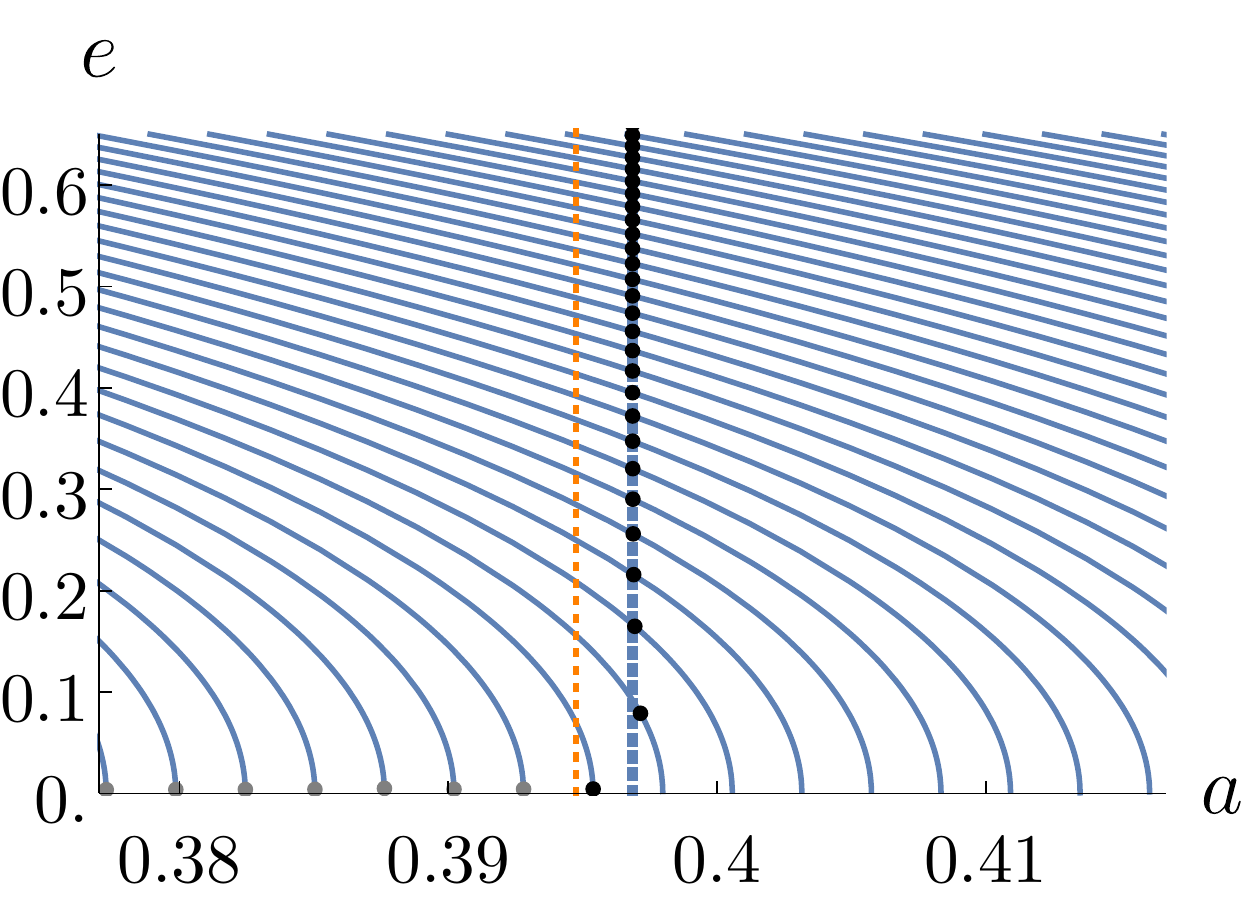}
\caption{$4:1$ resonance.}
\end{subfigure}

\caption{
Level curves of $\NN$ on the $(a,e)$ plane, for the case of the 2:1, 3:1  and 4:1 resonances respectively, 
with $a'=1~ \text{AU}$ for the perturber (in units where $\GravC M_*=1$). 
The solid lines depict equation \eqref{eq:aFromNNande}, with the numerical value of $\NN$ increases from left to right.
The vertical thick dashed lines indicate the location of exact Keplerian resonance, $a_{res}=a'(k/k')^{2/3}$.
The dots represent the equilibrium values for the eccentricity and the semi-major axis, on different level 
curves of $\NN$, while the arbitrary value of $\VARPI$ remains fixed. Here we used $e'=0.2$ and $\mu=10^{-3}$.
Notice the deviation of the equilibrium points from the exact resonance, which is particularly evident in the
case of first order resonances (the 2:1 resonance in this case), see the text for details. 
Since this deviation is linked to a faster precession of the pericenter $\VARPI=-\p$, 
the value of $e$ at which this effect becomes larger than $(2/3)(\sqrt{\mu}/k)(a_{res})^{5/2}$ 
yields a lower bound in $e$ above which our approach is valid. 
The orange dashed line indicates a deviation from the exact resonance of this amount.
We thus colour-coded the equilibrium points using black for those that fall above this lower limit in eccentricity, 
and gray for those that fall below it: 
for the latter, the fast change in $\p$ does not allow us to consider the pair $(\NN,\p)$ as slowly evolving variables.
}
\label{fig:Figure2} 
\end{figure}

We show an example of this calculation in Figure \ref{fig:Figure2}, for the 2:1, 3:1 and 4:1 resonances. 
It is worth pointing out that the equilibrium points in the $(e,a)$ diagram deviate away from the Keplerian 
resonant value $a_{res}=a' (k/k')^{2/3}$ as $e\to0$. 
This is especially evident in the case of first order resonances, $|k-k'| = 1$. 
In this case, for $e>e'$ the main harmonic in the Hamiltonian is $e\cos\SIGMA$, i.e.\ $\sqrt{\P}\cos\SIGMA$, from 
\eqref{eq:ModifDelaunayVariables} and expansion for small $e$. 
Because $\dot \p=\partial{\Ha}/\partial{\P}$, this harmonic gives $\dot\p\propto 1/\sqrt{\P}\sim 1/e$, 
which grows considerably as $e$ approaches zero;
therefore in order to maintain $\dot\SIGMA = [k'\dot\LAMBDA'-k\dot\LAMBDA+(k'-k)\dot\p]/(k'-k)\sim 0$ 
one must have $k'\dot\LAMBDA'-k\dot\LAMBDA \nsim 0$ i.e.\ $a/a' \nsim (k/k')^{(2/3)}$. 
For resonances of order $|k'-k| > 1$ the main harmonic in the Hamiltonian for $e>e'$ is $e^{|k'-k|}\cos\SIGMA$, 
i.e. $\P^{|k'-k|/2}\cos\SIGMA$. 
Therefore the first derivative in $\P$ is not singular for $e\sim\sqrt{\P}\to 0$. 
However, for $e<e'$ the main harmonic dependent on $e$ is $e'^{|k'-k|-1}e\cos[(k'-k)\SIGMA-(k'-k-1)(\p+\VARPI')]$, 
which gives a contribution to $\dot\p$ proportional to $1/e$, and the same reasoning applies.
Indeed, in the case of inner mean motion resonance, $a_{eq}$ always attains values that are slightly less 
than the Keplerian $a_{res}$ as $e\to0$, as shown in Figure \ref{fig:Figure2}.
We must note however that the deviation of the equilibrium points from the resonant value $a_{res}$
indicates a rapid precession of the pericenter $\VARPI$. 
This means that our assumption that $\p$ and $\NN$ remain constant is no longer valid. 
It breaks down when their motion evolves with a frequency of order $\sqrt{\mu}$, 
i.e. of the same order as the frequency of the $\SIGMAONA,\SIGMA$ evolution. 
When $\dot{\p}\sim\sqrt{\mu}$ the equilibrium semi major axis of the test particle deviates from the Keplerian value by 
the amount of order $(2/3)(\sqrt{\mu}/k) (a_{res})^{5/2}$. 
Thus, we can determine the lower limit in eccentricity for the validity of our approach as the value
of $e$ at which the equilibrium point $a_{eq}$ deviates away from $a_{res}$ by more than this quantity.
In this paper, we will focus mainly on resonances of order higher than 1, because they are much more 
efficient in pushing the eccentricity $e$ from $\sim 0$ to $\sim 1$ (see Section \ref{sec:Results}).
In this case, for $e<e'$ our approach is valid down to very small values of the eccentricity. 

We have implemented this scheme in a Mathematica notebook, available at \url{www.oca.eu/morby/SecResInMMR.nb}.
Let us now briefly explain the steps in our calculation, which ultimately yields the level curves of 
the Hamiltonian $\bar{\bar{\Ha}}(\NN,\p;J=0)$ on the $(e\cos\VARPI,e\sin\VARPI)$ plane,
thereby describing the secular evolution of the small body inside the mean
motion resonance with the perturber, in the limit of $J=0$.
First we consider a fixed value of $\NN=\NN^*$. 
For some given value of $\varpi$, we can find the (stable) equilibrium point in the $(e\cos\SIGMA, e\sin\SIGMA)$ 
plane in the following manner.
If $\VARPI=\VARPI_0=0$, the Hamiltonian \eqref{eq:AveragedHamiltonian} contains only cosines of $(k'-k)\SIGMA$ 
(and its multiples), so that its extrema in $\SIGMA$ are found at $2\pi l/(k'-k)$ and 
$(2\pi l + \pi)/(k'-k)$, $l\in\Z$. 
Taking e.g.\ $\SIGMA=\SIGMA_0=0,\pm\pi$ we can write the Hamiltonian \eqref{eq:AveragedHamiltonian} as 
$\bar\Ha(e, \NN^*,\SIGMA_0,\VARPI_0)$ and we can find its maximum as a function of the eccentricity.
The fact that $\bar{\Ha}$, as a function of $e$, must have a maximum at the resonance can be seen from 
\eqref{eq:AveragedKeplerianHamiltonian}, which clearly has a maximum in $\SIGMAONA=\SS-\NN$ at $\SIGMAONA_{res}$ 
(defined in \eqref{eq:LocationOfExactResonance}). 
Then, because $\frac{\partial^2 \Ha}{\partial\SIGMAONA^2}=\frac{\partial^2\Ha}{\partial\SS^2}$ and $\SS$ 
is monotonic in $e$, $\bar{\Ha}$ must have a maximum in $e$. 
Let's call $e_{max}$ the value of the eccentricity for which $\bar{\Ha}$ is maximal.
We must now check that this is in fact the stable equilibrium point, i.e.\ that 
in $\SIGMA_0$ the function $\bar\Ha(e_{max}, \NN^*,\SIGMA,\VARPI_0)$ of $\SIGMA$ has a maximum (and not a minimum).
If this is not the case, we can repeat the calculation with $\SIGMA=\pi/(k'-k), \pi/(k'-k)+\pi$. 
This effectively yields, for the given value of $\NN=\NN^*$ and for $\VARPI=\VARPI_0=0$, the stable equilibrium 
point in $(e,\SIGMA)$, denoted by $(e_{eq}, \SIGMA_{eq})$. 
Notice from Figure \ref{fig:Figure1} that $e_{eq}$ increases with the value $\NN^*$.

Following the procedure described above, we can assign to the Hamiltonian $\bar{\bar{\Ha}}(\NN^*,\VARPI_0;J=0)$ the value 
$\bar\Ha(e_{eq}, \NN^*,\SIGMA_{eq}, \VARPI_0)$ on the point $(e_{eq} \cos\VARPI_0, e_{eq} \sin\VARPI_0)$.
Note also that from the equilibrium value $e_{eq}$, one can obtain the corresponding $a_{eq}$ 
through equation \eqref{eq:aFromNNande}. 
When $\VARPI$ is not zero, the diagram in the $(e\cos\SIGMA, e\sin\SIGMA)$ plane is, to a good approximation\footnote{
We have checked that in the 4:1 resonance $e_{eq}$ changes only by $\lesssim 0.1$\% with the rotation of $\VARPI$, 
down to $e_{eq}\sim0.05$, for $e'=0.1$.
}
for most values of $e$, simply rotated by 
a quantity related to $\VARPI$, so that the equilibrium values of the eccentricity and the semi-major axis don't 
change substantially, but only $\SIGMA_{eq}$ changes (see Figure \ref{fig:Figure1}). 
This way, one can obtain the equilibrium value for $\SIGMA$ by finding the maximum in $\SIGMA$ of the function 
$\bar\Ha(e_{eq}, \NN^*, \SIGMA,\VARPI)$, for the fixed value of $\VARPI$. 
It is worth noticing that in order to assign to the point $(e_{eq}\cos\VARPI, e_{eq}\sin\VARPI)$ the appropriate 
value of the Hamiltonian, we are only interested in the
$\max_{\SIGMA\in[0,2\pi]}\bar\Ha(e_{eq}, \NN^*, \SIGMA,\VARPI)=
\max_{\SIGMA\in[0,2\pi/(k'-k)]}\bar\Ha(e_{eq}, \NN^*, \SIGMA,\VARPI)$ for the fixed value of $\VARPI$,
not on the actual value $\SIGMA_{eq}$ of $\SIGMA$ where the maximum is attained. 

By letting $\NN$ vary, i.e.\ effectively by allowing $e_{eq}$ to vary, we obtain the level curves of the
Hamiltonian $\bar{\bar\Ha} (\NN,\p;J=0)$ in the variables $(e\cos\VARPI,e\sin\VARPI)$. 
We present several examples in Section \ref{sec:Results}, where we show level curves of $\bar{\bar\Ha}$ 
in for different resonances and different values of $e'$.

\section{The effect of short-range forces}\label{sec:GRContribution}

When the eccentricity of the test mass reaches values close to 1, so that the osculating ellipse becomes narrower and narrower,
the periapsis distance from the star $a_{peri}=a(1-e)$ becomes considerably small. 
At this point, the effect of various short-range forces may become important and must be considered.
One such short-range force arises from General Relativity, with the post-Newtonian contribution to the test particle's 
Hamiltonian given by
\begin{equation}
\Ha_{GR} = \frac{\GravC M_*}{a} \left(\frac{\GravC M_*}{a c^2}\right) \left(\frac{15}{8}-\frac{3}{\sqrt{1-e^2}}\right),
\end{equation}
where $c$ is the speed of light (\cite{1986SvA....30..224K}).
Note that the 15/8 term gives the General Relativity correction to the mean motions only, while the $1/\sqrt{1-e^2}$ term
gives the correction to the precession of the pericenter. Since we are interested only in the latter and we have averaged
over the mean motion, we drop the former. 
Another short-range force arises 
from the rotational bulge of the central star, with the Hamiltonian given by 
\begin{equation}
{\Ha}_{rot}=-{\GravC M_* R_*^2 J_2\over 2 a^3 (1-e^2)^{3/2}},
\end{equation}
where $R_*$ is the stellar radius, and $M_* R_*^2 J_2$ is the rotation-induced quadruple moment of the star. 
To assess the importance of these short-range forces, we compare ${\Ha}_{GR}$ and ${\Ha}_{rot}$ to $\Phi_0$, the characteristic 
tidal potential produced by the planetary perturber on the test particle,
\begin{equation}
\Phi_0\equiv {\GravC m' a^2\over {a'}^3}.
\end{equation}
We find
\begin{equation}
\begin{split}
{|{\cal H}_{GR}|\over \Phi_0}&\simeq  
10^{-2}\left(\!{M_*\over M_\odot}\!\right)^2\left({m'\over M_\oplus}\right)^{\!-1}
\left({a'\over a}\right)^3 \times \\
 &\quad \times \left({a\over{\rm AU}}\right)^{\!-1}\!{1\over (1-e^2)^{1/2}}\\
&\simeq  1.7 \left(\!{M_*\over M_\odot}\!\right)^2\left({m'\over M_\oplus}\right)^{\!-1}
\left({n\over 4n'}\right)^2 \times  \\
& \quad \times \left({a\over{\rm AU}}\right)^{\!-1/2}\!
\left({a_{\rm peri}\over R_\odot}\right)^{\!-1/2},
\end{split}
\end{equation}
\begin{equation}
\begin{split}
{|{\cal H}_{\rm rot}|\over \Phi_0}&\simeq {k_{q*}{\hat\Omega_*}^2\over 2}
\left({M_*\over m'}\right)\left({R_*\over a}\right)^2\!
\left({a'\over a}\right)^3\!{1\over (1-e^2)^{3/2}}\\
&\simeq 0.086 \left({k_{q*}\over 0.01}\right)
\left(\!{P_*\over 10\,{\rm day}}\!\right)^{\!-2}
\left({R_*\over R_\odot}\right)^5
\left({m'\over M_\oplus}\right)^{\!-1} \times \\
&\quad \times\left({n\over 4n'}\right)^2\left({a\over{\rm AU}}\right)^{\!-1/2}\! 
\left({a_{\rm peri}\over R_\odot}\right)^{\!-3/2}, 
\end{split}
\end{equation}
where $a_{peri}=a(1-e)$, and we have used $J_2=k_{q*}{\hat\Omega_*}^2=k_{q*}\Omega_*^2 R_*^3/(GM_*)$, with 
$\Omega_*=2\pi/P_*$ the stellar rotation rate.
Clearly, for most main-sequence stars (with $P_*\gtrsim 2$~days) and white dwarfs, $|{\cal H}_{\rm rot}|$ is negligible 
compared to $|{\cal H}_{\rm GR}|$.
We will neglect ${\cal H}_{\rm rot}$ in the remainder of this paper.

It is straightforward to include $\Ha_{GR}$ into the scheme outlined in Section \ref{sec:AdiabaticMethod}, as we simply need to 
add the value $\Ha_{GR}(a_{eq}, e_{eq})$ to the value of the planetary Hamiltonian. 
This could change the dynamics of the system considerably at sufficiently high eccentricity. 
In fact, up to this point, the perturber's mass (rescaled by the star's mass) $\mu$ has played a role in 
setting the amplitude of libration in $\SIGMAONA$ (or $a$, $e$) in the Hamiltonian $\bar{\Ha}_{(\NN,\p)}$, 
as well as the frequency of libration around the stable equilibrium point.  
However, the dynamics described by $\bar{\bar{\Ha}}(\NN,\p;J=0)$ does not depend on the perturber's mass, 
because both $\NN$ and $\p$ appear only in the part of the Hamiltonian derived from $\bar{\Ha}_{pert}$, 
where $\mu$ is a multiplying parameter. 
Thus, the evolution of $e$ as a function of $\VARPI=-\p$ does not depend on $\mu$. 
Only the timescale of this evolution occurs does (and scales as $1/\mu$).  

With the addition of the General Relativity term in the Hamiltonian, the dynamical behavior of the system will 
in general depend on $\mu$. 
Indeed, 
$\Ha_{GR}$ is independent of $\mu$, and is dependent on $\NN$:
\begin{equation}
\Ha_{GR}(\SIGMAONA,\NN,\SIGMA,\p)=\frac{3 \GravC^4 M_*^4}{c^2} \frac{(k'-k)^4}{k^3 \SIGMAONA^3 (k \NN - (\NN + \SIGMAONA) k')}.
\end{equation}
Thus, the actual evolution of $\NN$ (i.e.\ of $e_{eq}$ if $J=0$) as a function of $\p$ (i.e.\ $\VARPI$) depends on the value 
of $\mu$. 

Another way to understand this is that the General Relativity potential has the effect of keeping the eccentricity 
constant while the pericenter $\VARPI=-\p$ precesses 
(because $\dot\p=\frac{\partial \Ha_{GR}}{\partial \NN} =
-\frac{3 \GravC^4 M_*^4}{c^2} \frac{(k'-k)^5}{k^3 \SIGMAONA^3 (k \NN - (\NN + \SIGMAONA) k')^2} <0$
while $\dot\NN=0$, $\dot\SIGMAONA=0$).
In contrast, in the restricted three-body problem (see previous section) the precession of the pericenter 
is coupled with the variation in the eccentricity. 
Since the mass of the perturber $\mu$ appears in the planetary potential as a multiplicative factor in the perturbation,
but not in the General Relativity potential, it will play the role of a parameter regulating which of the two
dynamics in the $(e\cos\VARPI,e\sin\VARPI)$ plane is dominant. 
The smaller $\mu$ is, the more the General Relativity contribution will be 
prevailing, and the less efficient the planet will be in pumping the eccentricity of the test particle; the bigger $\mu$ is,
the less the $\Ha_{GR}$ contribution will be apparent.

\section{Results}\label{sec:Results}

\begin{figure*}
\centering
\begin{subfigure}[b]{0.45 \textwidth}
\centering
\includegraphics[scale=0.65
]{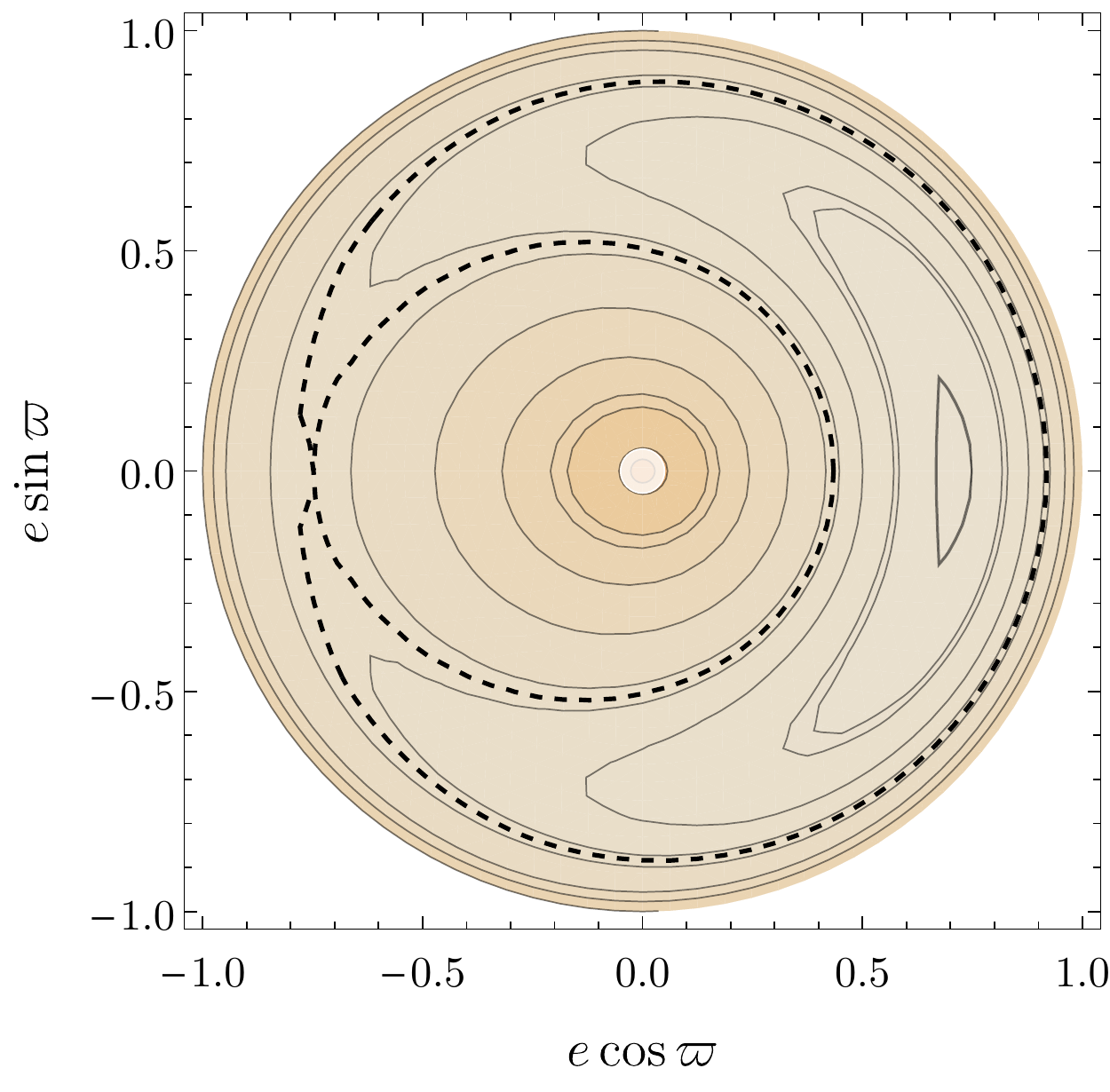}
\caption{$e'=0.05$.}
\end{subfigure}
\begin{subfigure}[b]{0.45 \textwidth}
\centering
\includegraphics[scale=0.65
]{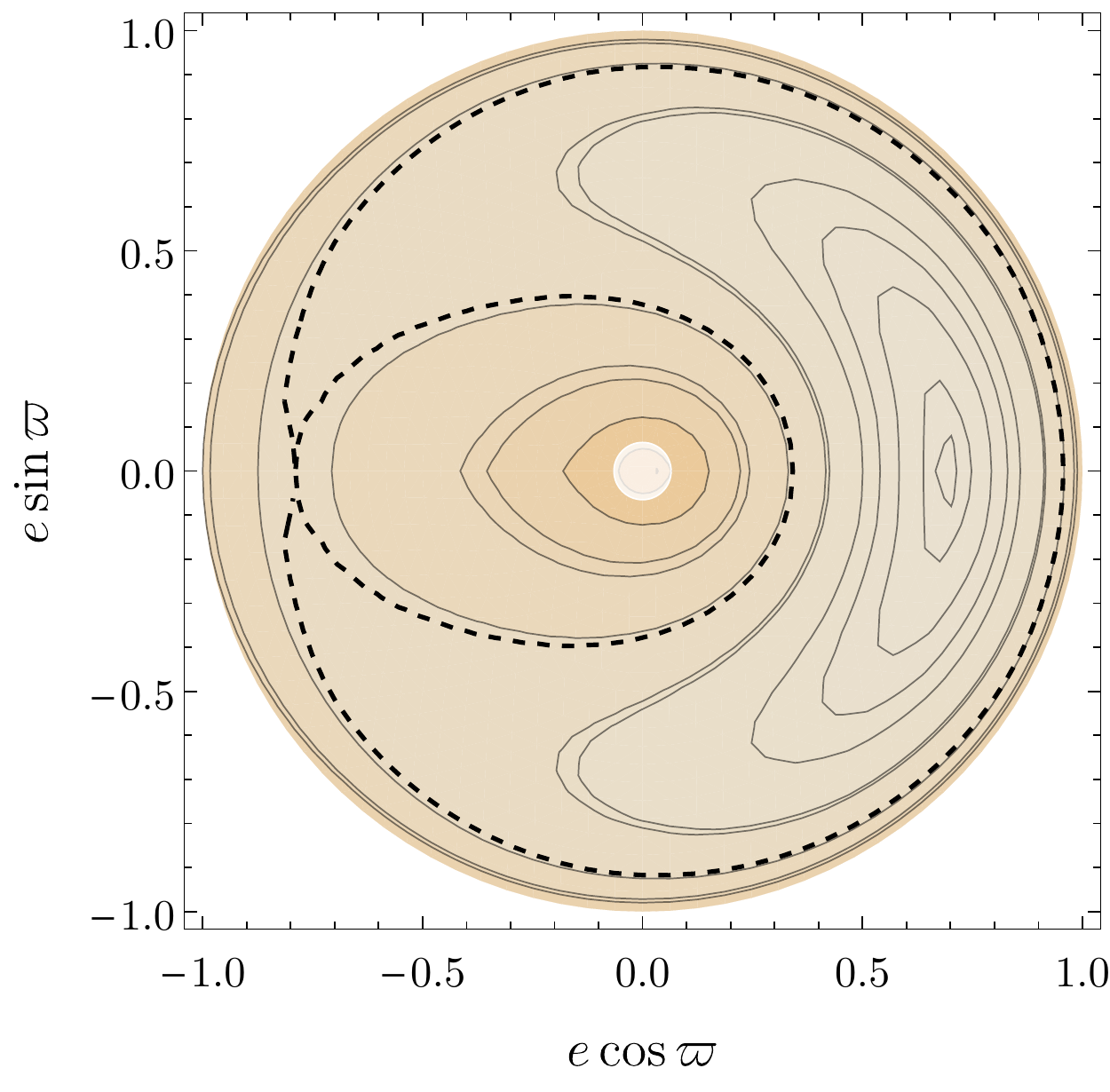}
\caption{$e'=0.1$.}
\end{subfigure}

\caption{Level curves of the Hamiltonian $\bar{\bar\Ha}$ on the $(e\cos\VARPI, e\sin\VARPI)$ plane for the 2:1 mean motion
resonance with an outer perturber,
for low values of $e'$ (=0.05 and 0.1, both with $\VARPI'=0$). 
The General Relativity effect is not included.
Lighter colours indicate a higher value of the Hamiltonian.
The white shaded disks centered at the origin indicate the regions where the adiabatic method is not valid 
(see Section \ref{sec:AdiabaticMethod}), i.e.\ where our calculations do not necessarily reflect the true dynamics 
of the system.
The dark dashed line indicates a set of critical orbits which separate the phase space into a circulation zone near the origin,
a libration zone near the stable equilibrium point at $\VARPI=0$, and an outer circulation region.
All orbits with initially low eccentricities do not experience appreciable increase in $e$.   
}
\label{fig:Figure3} 
\end{figure*}

\begin{figure*}
\centering
\begin{subfigure}[b]{0.45 \textwidth}
\centering
\includegraphics[scale=0.65
]{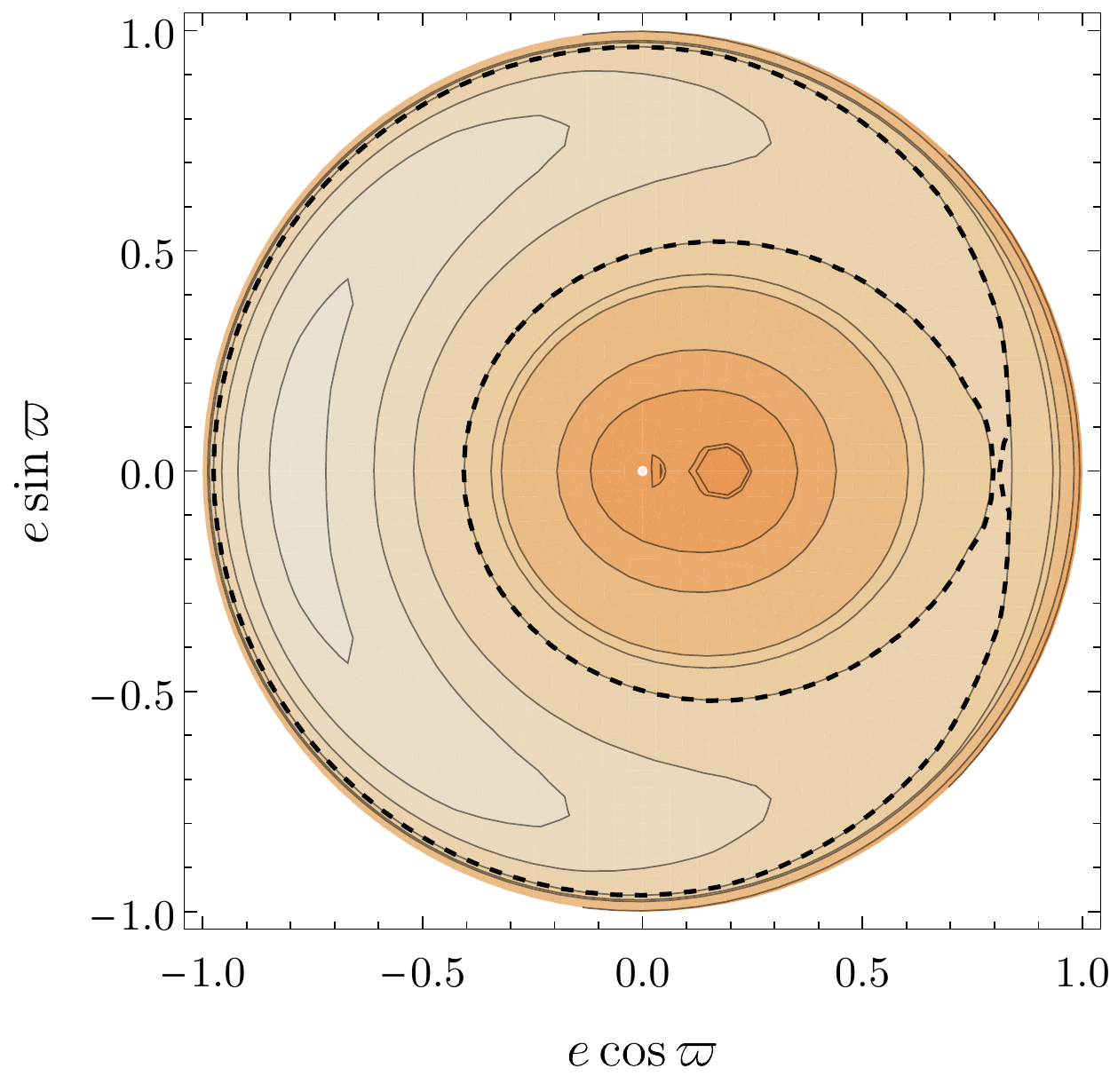}
\caption{$e'=0.05$.}
\end{subfigure}
\begin{subfigure}[b]{0.45 \textwidth}
\centering
\includegraphics[scale=0.65
]{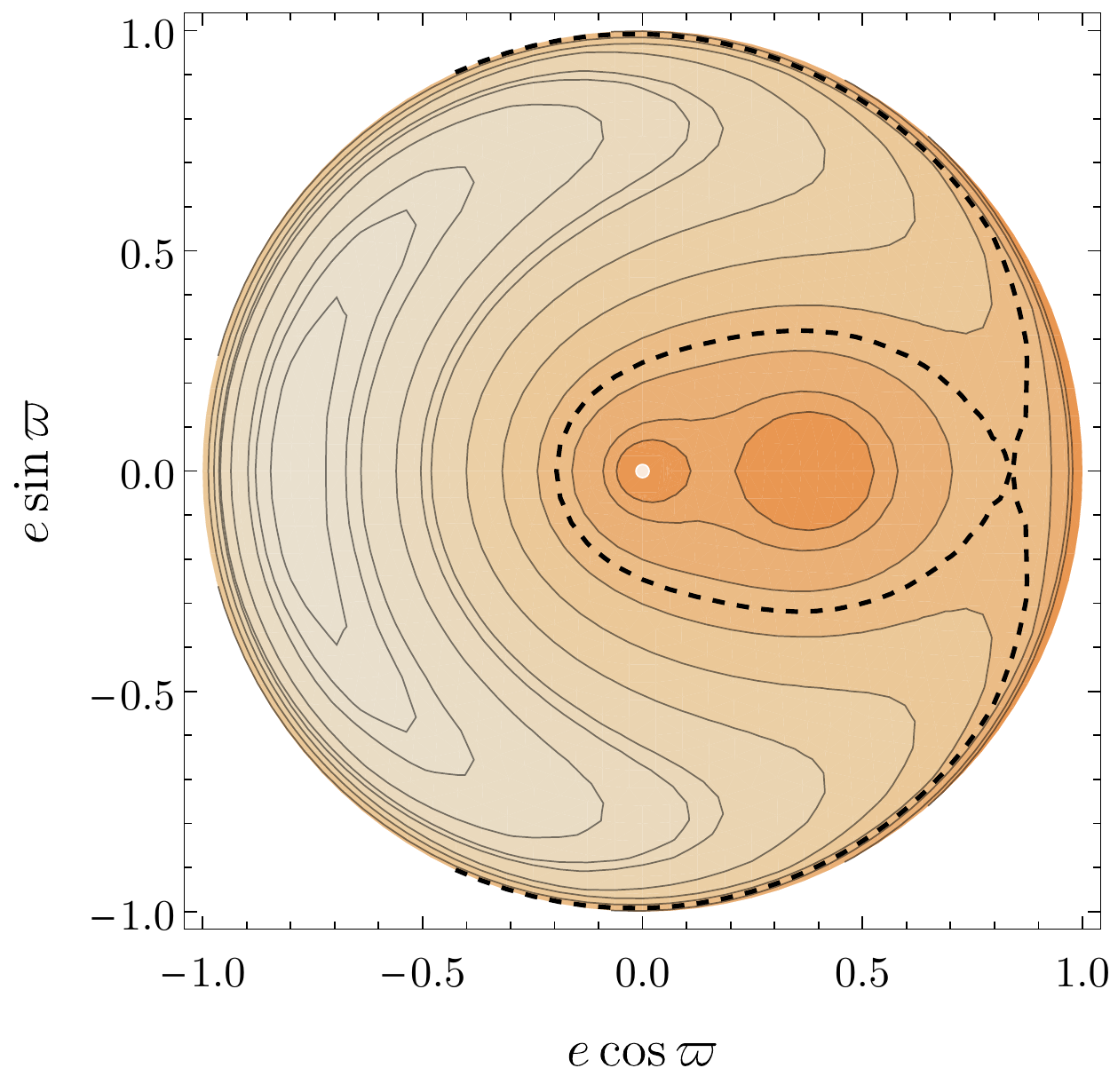}
\caption{$e'=0.1$.}
\end{subfigure}

\caption{Same as Figure \ref{fig:Figure3}, for the 3:1 mean motion resonance with an outer perturber.
}
\label{fig:Figure4} 
\end{figure*}

\begin{figure*}
\centering
\begin{subfigure}[b]{0.45 \textwidth}
\centering
\includegraphics[scale=0.65
]{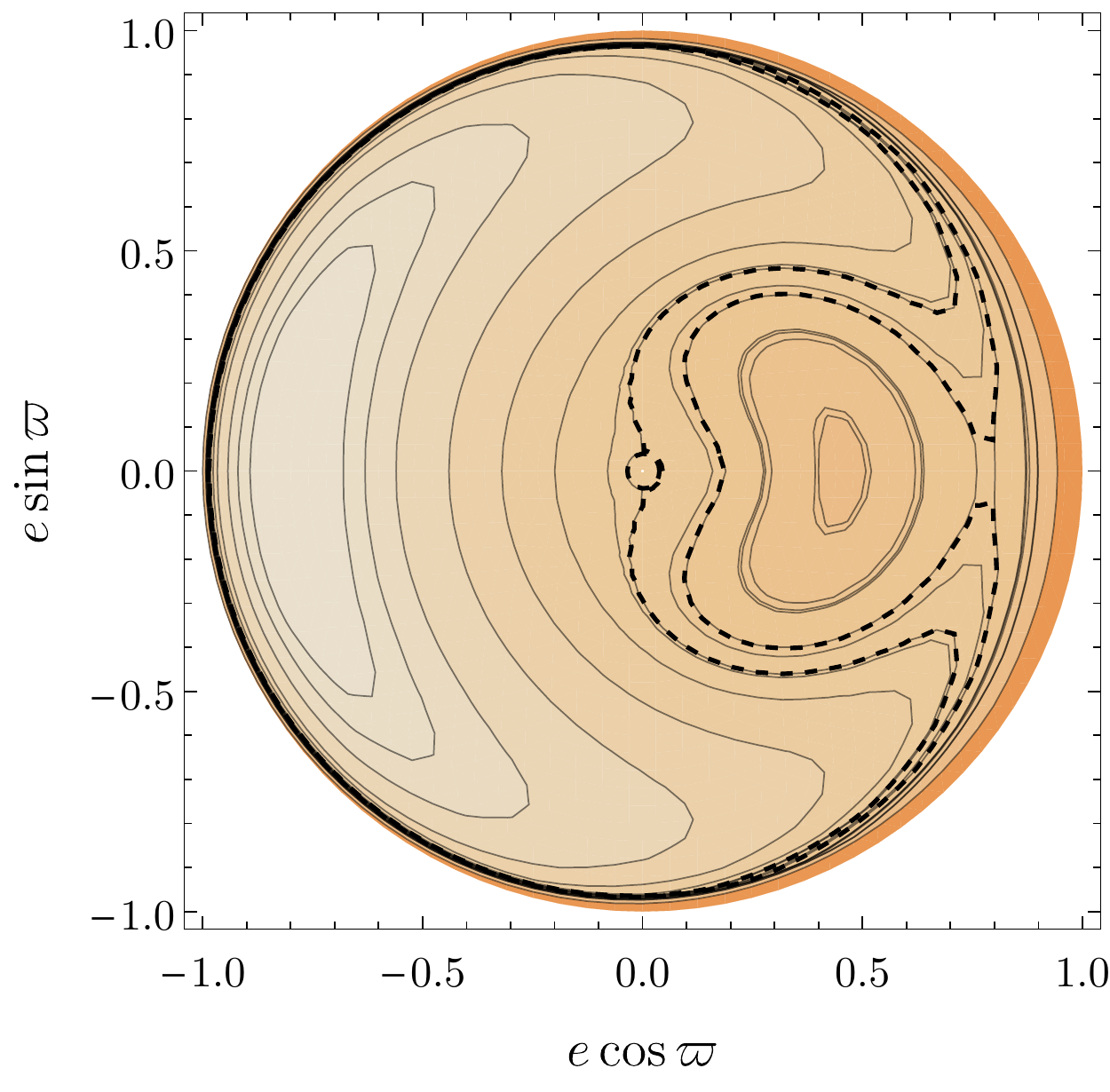}
\caption{$e'=0.05$.}
\end{subfigure}
\begin{subfigure}[b]{0.45 \textwidth}
\centering
\includegraphics[scale=0.65
]{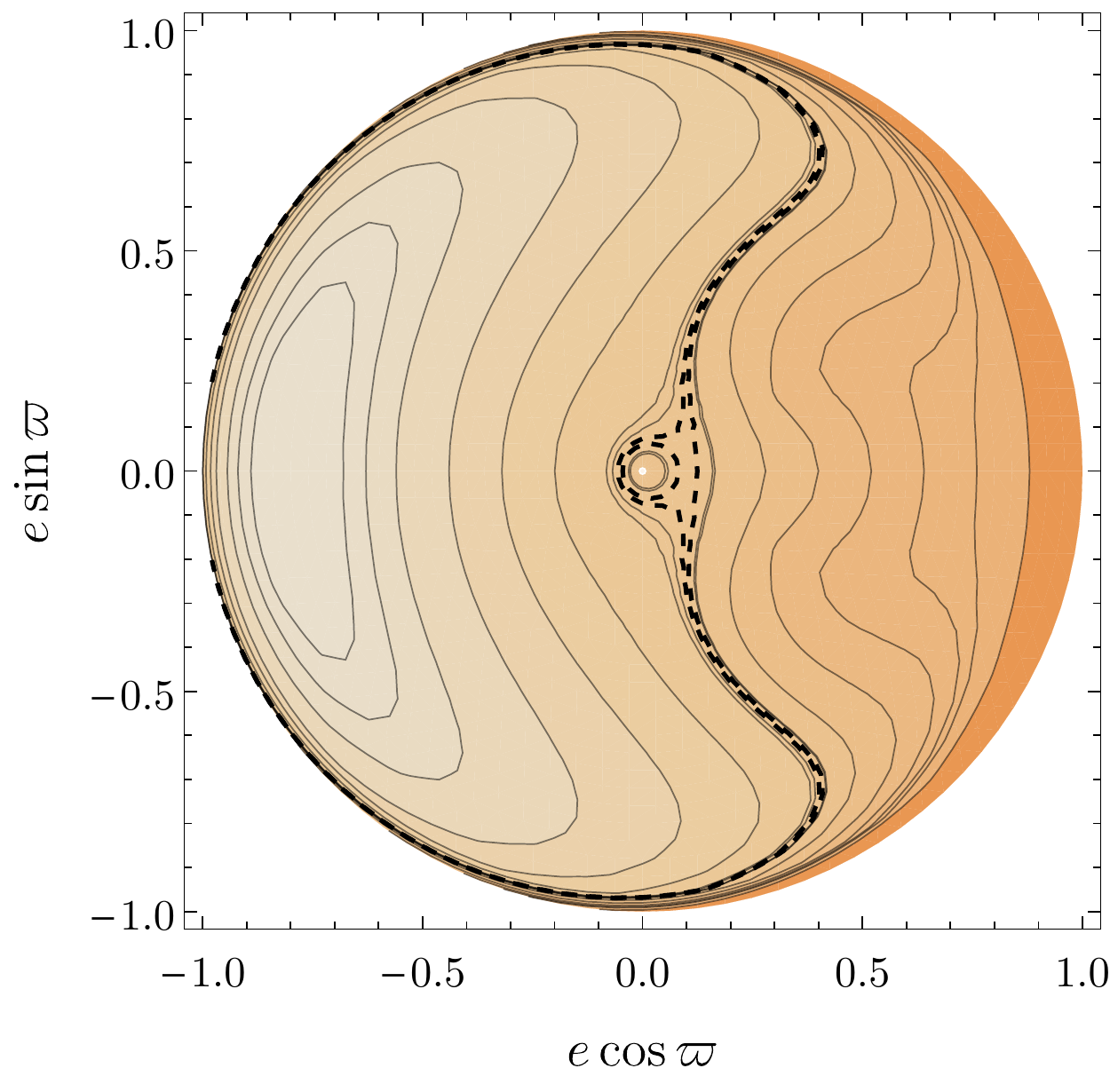}
\caption{$e'=0.1$.}
\end{subfigure}

\caption{Same as Figure \ref{fig:Figure3}, for the 4:1 mean motion resonance with an outer perturber.
In contrast to Figures \ref{fig:Figure3} and \ref{fig:Figure4}, orbits with small initial eccentricities can be driven to $e\sim1$.
}
\label{fig:Figure5} 
\end{figure*}

In Figures \ref{fig:Figure3}, \ref{fig:Figure4} and \ref{fig:Figure5} 
we show the level curves of the Hamiltonian $\bar{\bar\Ha}$ (see Section \ref{sec:AdiabaticMethod}) 
on the $(e\cos\VARPI, e\sin\VARPI)$ plane for the 2:1, 3:1 and 4:1
resonances respectively, with low values of the eccentricity of the perturber, $e'=0.05$, and $e'=0.1$. 
The General relativity effect is not included in these figures.
The white shaded disks centered at the origin (barely visible in Figure \ref{fig:Figure5}) 
indicate the regions where the adiabatic method is not valid (see section \ref{sec:AdiabaticMethod}): 
in these regions our calculations do not necessarily reflect the true dynamics 
of the system.

Note from Figure \ref{fig:Figure5} how even for small values of $e'$ the 4:1 resonance is extremely effective in 
driving the eccentricity of the test particle from $e\sim0$ to $e\sim1$. 
Indeed, there is only a small portion of the phase space that allows orbits starting with low-eccentricities 
to circulate near the origin ($e=0$). 
In the case of $e'=0.05$ only some orbits with moderate initial eccentricities, i.e.\ $e>0.2$ and initial $\VARPI\sim0$, 
actually librate around the stable equilibrium point at $\VARPI=0$, $e\sim0.4$, 
while for $e'=0.1$, all orbits sufficiently distant from the origin eventually end up at $e\sim1$.  
This is not the case for the other resonances. 
For the 2:1 resonance, we see from Figure \ref{fig:Figure3} that all orbits with initial eccentricities up to $\sim 0.4$ and 
$\sim 0.3$, for $e'=0.05$ and $e'=0.1$ respectively, remain confined around the equilibrium point near the origin. 
Another equilibrium point is present at $e\sim0.7$, $\VARPI=0$, implying that whatever the initial values of $\VARPI$ even a higher
initial eccentricity is not enough to push the test particle to a star-grazing orbit.
Indeed, the presence of the separatrix (shown as a black dashed curve) does not allow any orbit with initial eccentricity
lower than $\sim 0.9$ to move farther away from the origin.
In the case of the 3:1 resonance, Figure \ref{fig:Figure4} shows that eccentricities smaller than 0.4 for $e'=0.05$ 
and 0.2 for $e'=0.1$ remain small, as the level curves librate around $\VARPI=0$.
For $e'=0.05$ a separatrix bounds the maximal attainable eccentricity as in the 2:1 resonance. 
This confirms the results in \cite{1996Icar..120..358B}.

\begin{figure*}
\centering
\begin{subfigure}[b]{0.45 \textwidth}
\centering
\includegraphics[scale=0.65
]{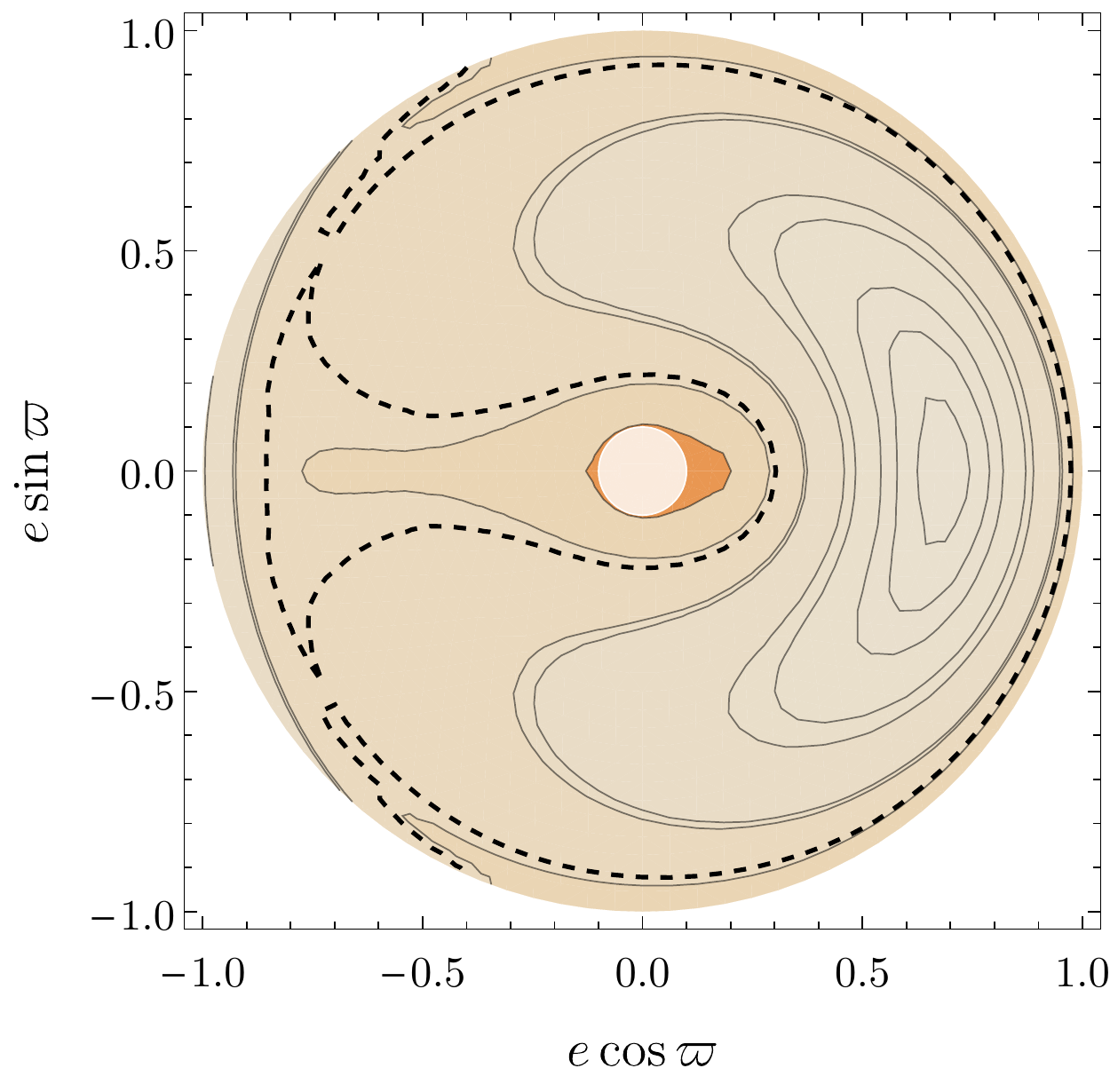}
\caption{$e'=0.2$.}
\end{subfigure}
\begin{subfigure}[b]{0.45 \textwidth}
\centering
\includegraphics[scale=0.65
]{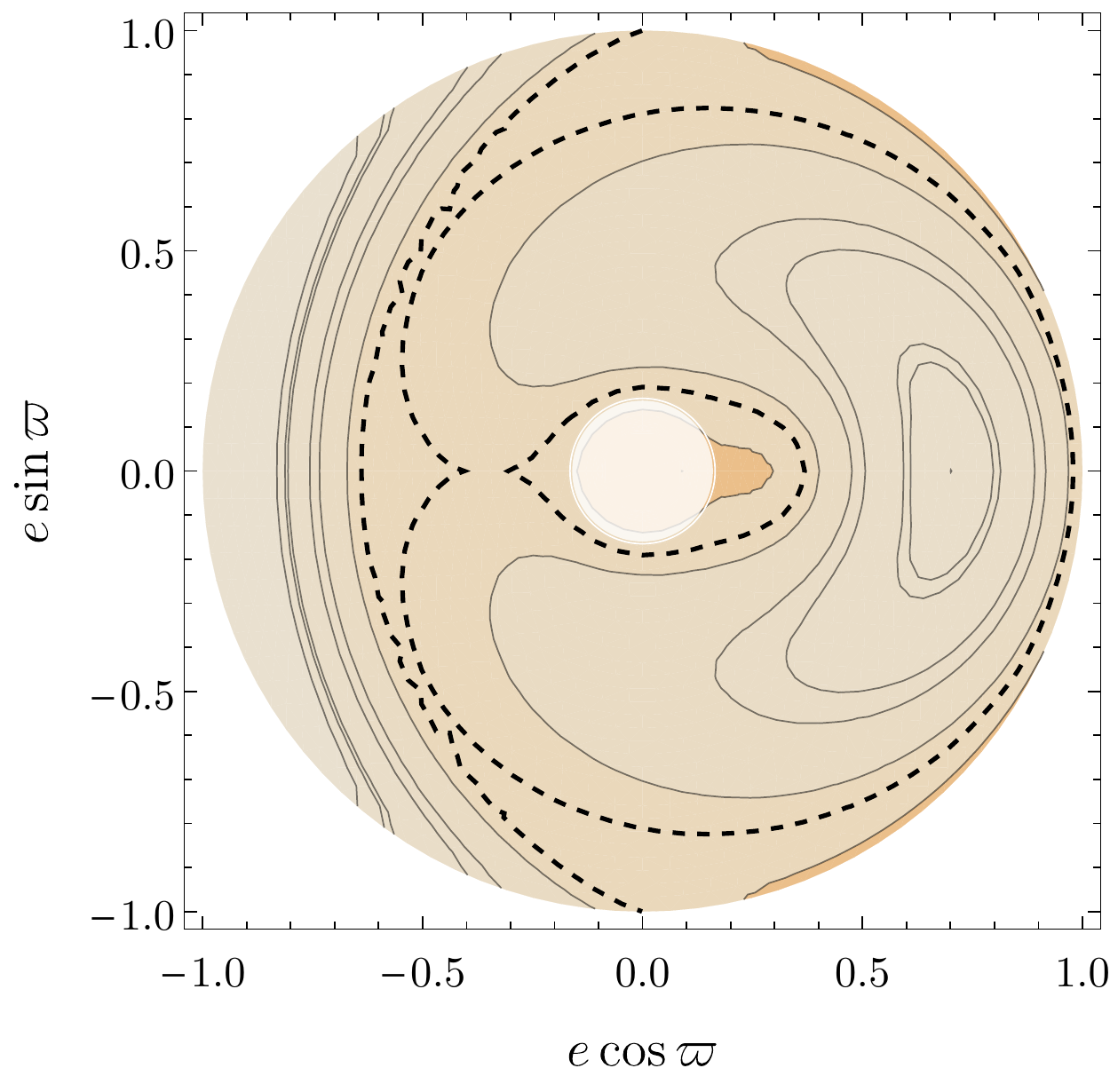}
\caption{$e'=0.3$.}
\end{subfigure}

\caption{Level plots of the Hamiltonian $\bar{\bar\Ha}$ on the $(e\cos\VARPI, e\sin\VARPI)$ plane for the 2:1 mean motion
resonance with an outer perturber with modest eccentricities ($e'=0.2$ and $e'=0.3$, both with $\VARPI'=0$).
All orbits with initially low eccentricities do not experience a large increase in $e$.
}
\label{fig:Figure6} 
\end{figure*}

\begin{figure*}
\centering
\begin{subfigure}[b]{0.45 \textwidth}
\centering
\includegraphics[scale=0.65
]{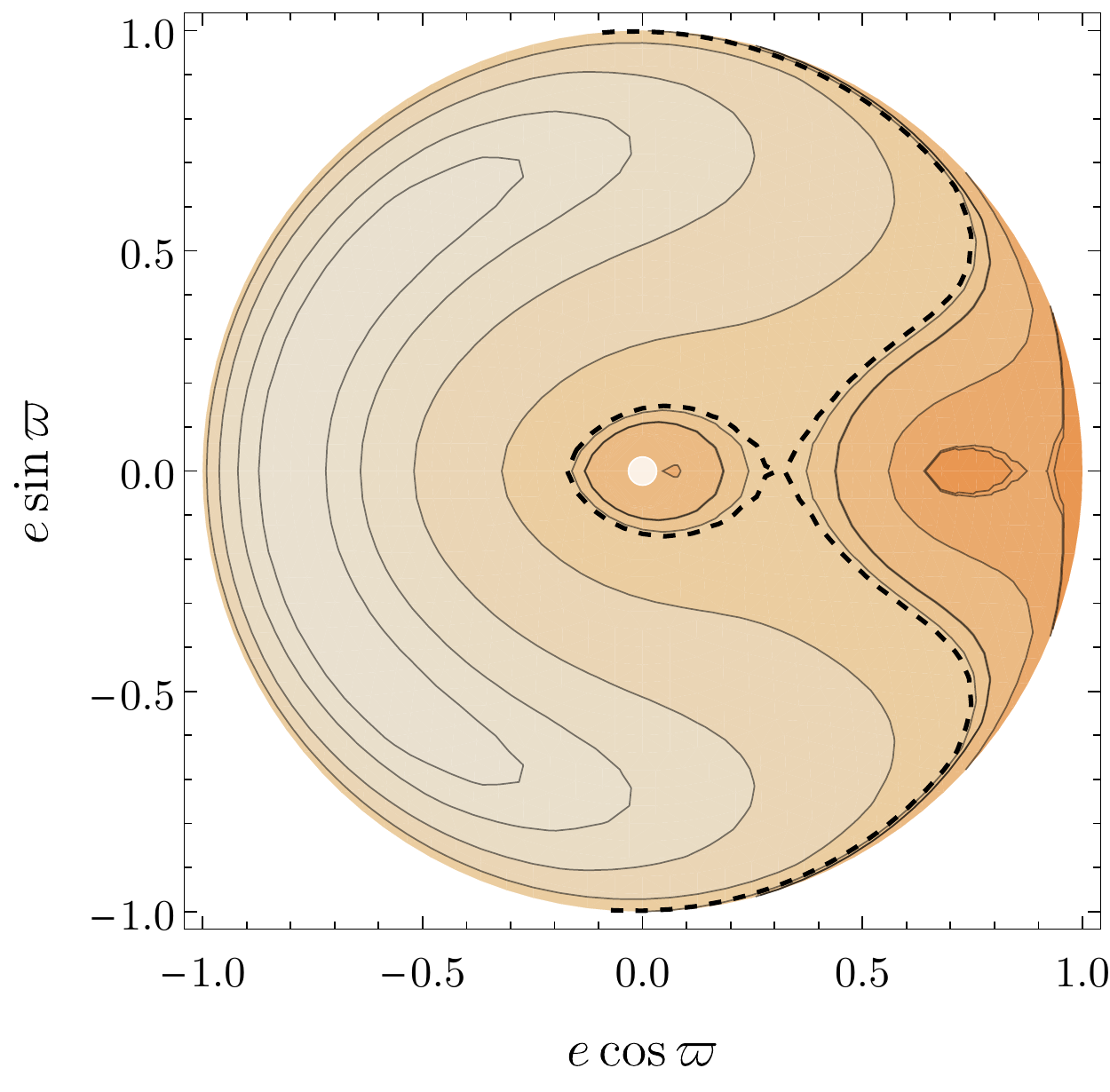}
\caption{$e'=0.2$.}
\end{subfigure}
\begin{subfigure}[b]{0.45 \textwidth}
\centering
\includegraphics[scale=0.65
]{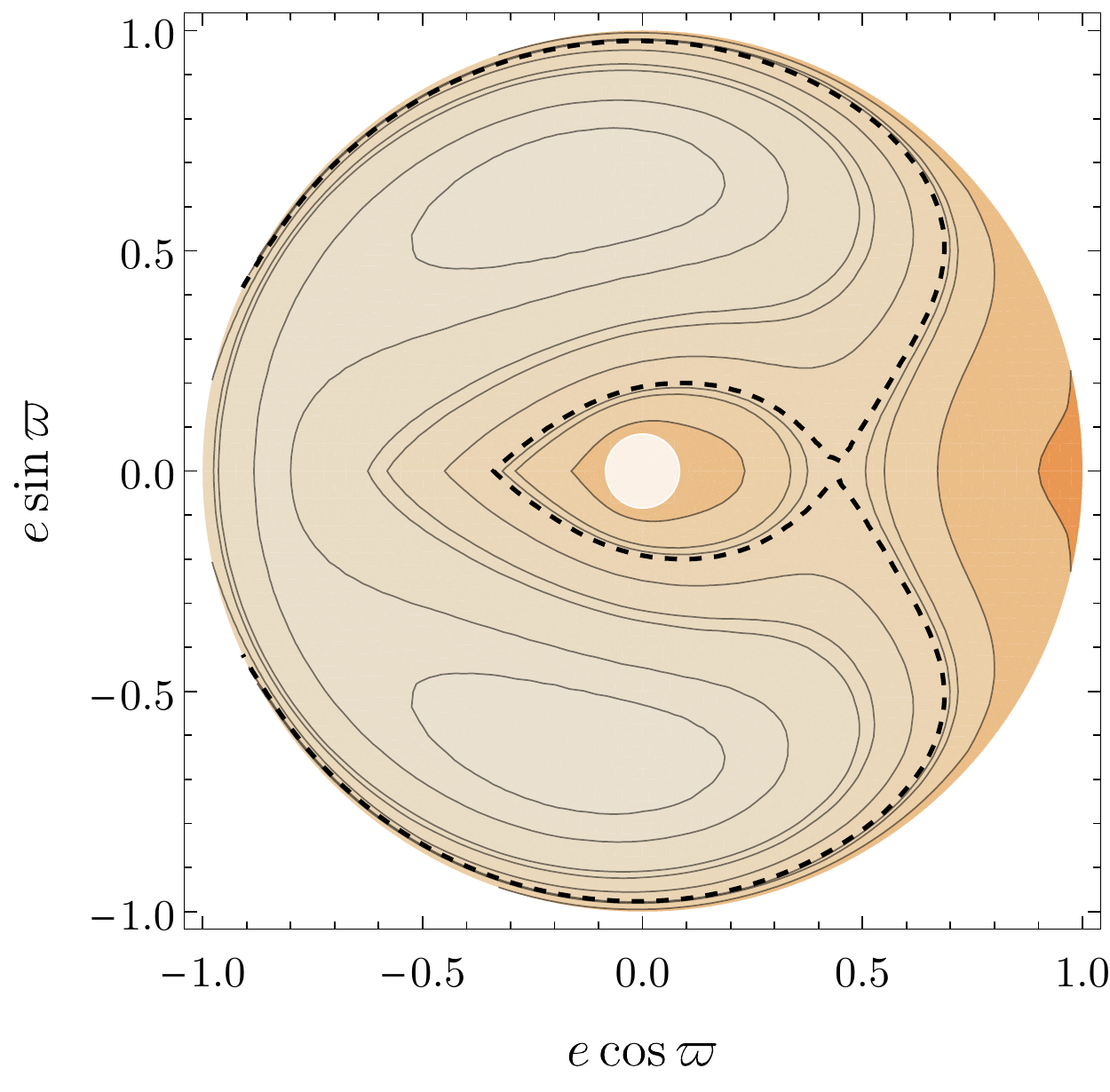}
\caption{$e'=0.3$.}
\end{subfigure}

\caption{Same as Figure \ref{fig:Figure6}, for the 3:1 mean motion resonance with an outer perturber.
}
\label{fig:Figure7} 
\end{figure*}

\begin{figure*}
\centering
\begin{subfigure}[b]{0.45 \textwidth}
\centering
\includegraphics[scale=0.65
]{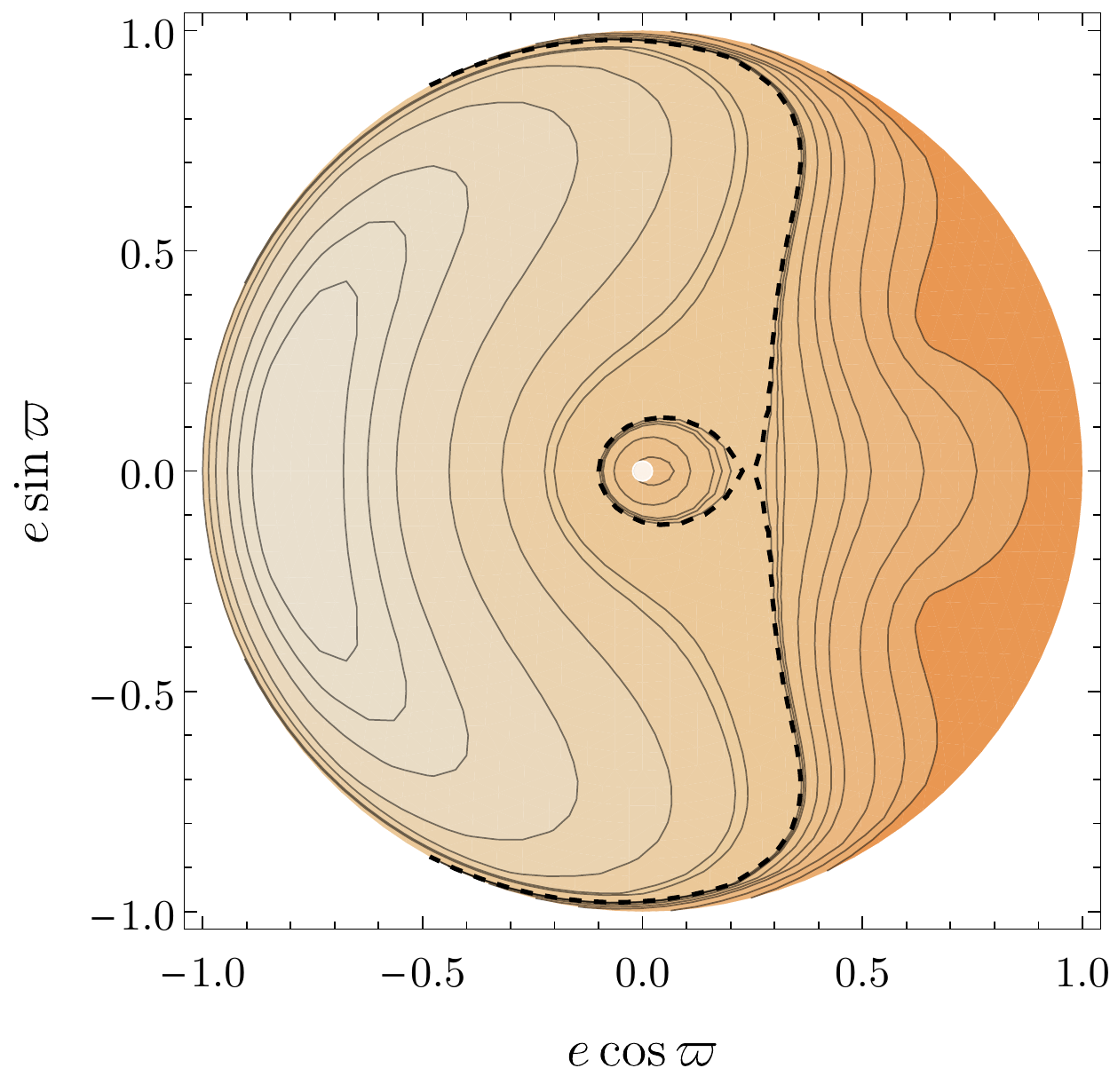}
\caption{$e'=0.2$.}
\end{subfigure}
\begin{subfigure}[b]{0.45 \textwidth}
\centering
\includegraphics[scale=0.65
]{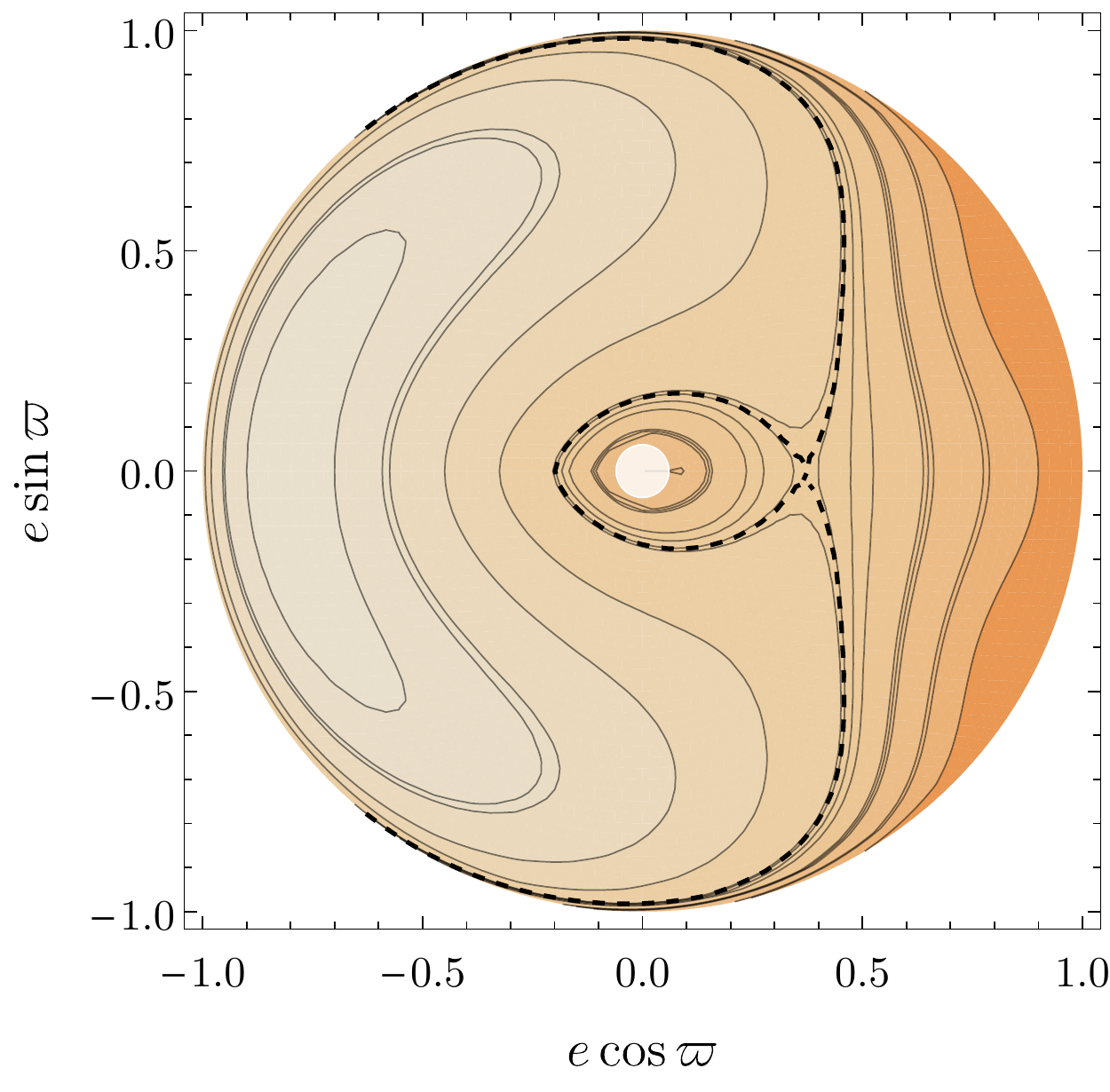}
\caption{$e'=0.3$.}
\end{subfigure}

\caption{Same as Figure \ref{fig:Figure6}, for the 4:1 mean motion resonance with an outer perturber.
In contrast to Figure \ref{fig:Figure5}, orbits with initial $e\sim0$ do not experience extreme eccentricity excitation.
}
\label{fig:Figure8} 
\end{figure*}

Figures \ref{fig:Figure6}, \ref{fig:Figure7} and \ref{fig:Figure8} depict our results for high values of 
the perturber's eccentricity, $e'=0.2$ and $e'=0.3$.
We find that for the 2:1 and 3:1 resonances, even these higher values of $e'$ are not sufficient to generate 
star-grazing objects from $e\sim 0$.
Although one can see that for some initial configurations it is possible to observe an excitation in the eccentricity 
(see for example the case of the 3:1 resonance in Figure \ref{fig:Figure7}(a), where particles with $e\sim 0.2$ and $\VARPI=\pi$ 
may indeed reach $e\sim 1$), a modest/high initial eccentricity of the test particle is needed in order to eventually reach
a value close to 1. 
On the other hand, Figure \ref{fig:Figure8} shows that when the perturber's eccentricity is too large, 
the capability of the 4:1 resonance to raise the eccentricity of the test particle from $\sim0$ to $\sim1$ is diminished.
In all cases (Figures \ref{fig:Figure6}-\ref{fig:Figure8}), a separatrix confines all orbits close to the origin. 
Note that this separatrix occupies the region where the adiabatic method remains valid (see Section \ref{sec:AdiabaticMethod}), 
i.e.\ outside the white shaded region in each plot.
Therefore any orbit with a small initial eccentricity remains confined to low values of $e$.

As we noted in Section \ref{sec:GRContribution}, when the eccentricity of the test particle reaches sufficiently high values, 
the effect of the General Relativity term becomes
important, and the mass parameter $\mu$ plays a crucial role in shaping the dynamics of the system.
Led by our results shown in Figures \ref{fig:Figure3}-\ref{fig:Figure8}, we restrict ourselves to the case 
of a test particle in the 4:1 mean motion resonance with the
outer perturber, and we study the critical value $\mu_{crit}$ needed so that the periapsis distance $a_{peri}=a(1-e)$ 
reaches sufficiently small values, e.g.\ the radius of the central star or the star's Roche limit
(which is $\sim R_\odot$, for white dwarfs and asteroids with internal density about a few ${\rm g}/{\rm cm}^3$).

\begin{figure*}
\centering
\begin{subfigure}[b]{0.45 \textwidth}
\centering
\includegraphics[scale=0.65
]{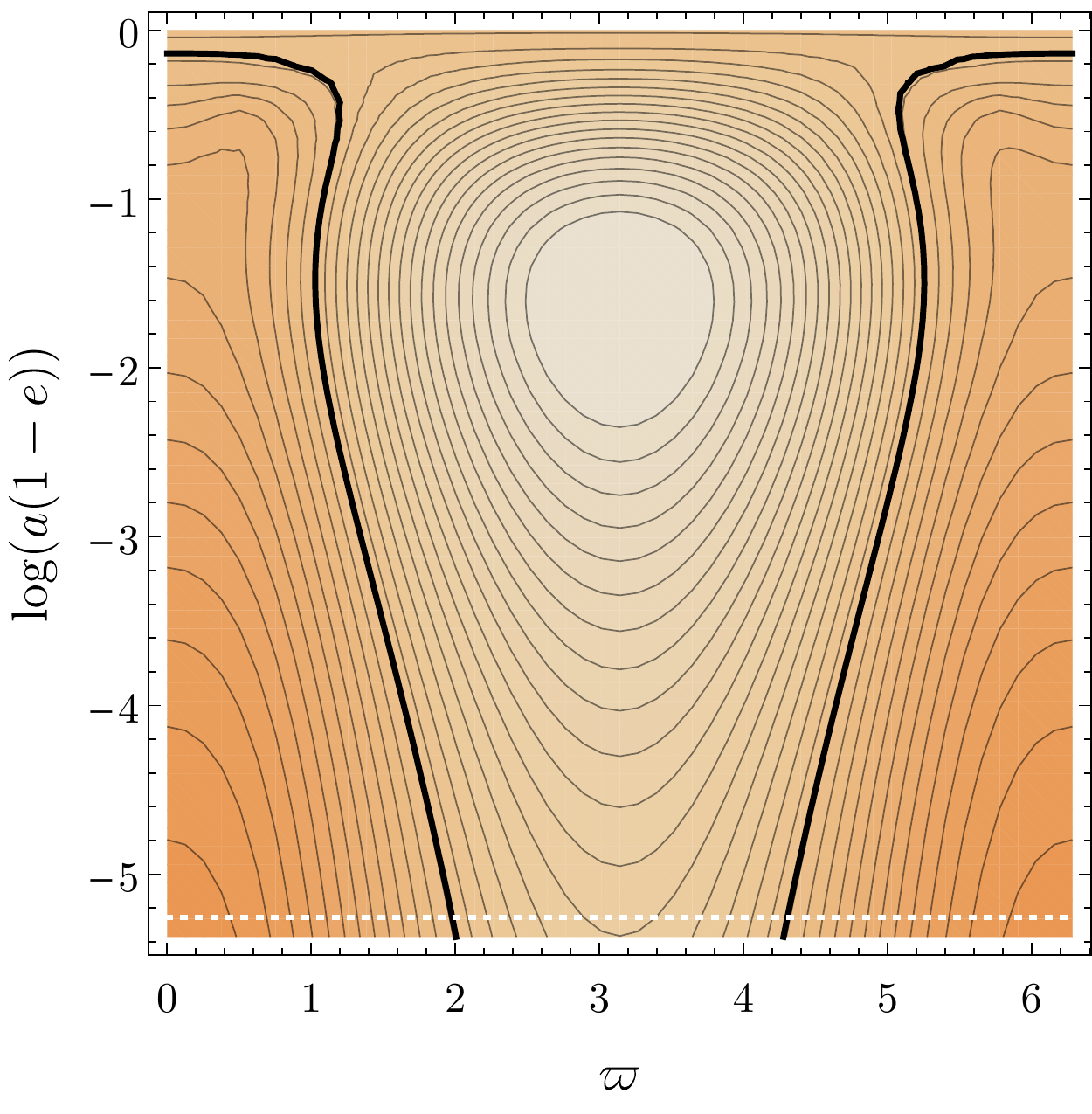}
\caption{Purely planetary case.}
\end{subfigure}
\begin{subfigure}[b]{0.45 \textwidth}
\centering
\includegraphics[scale=0.65
]{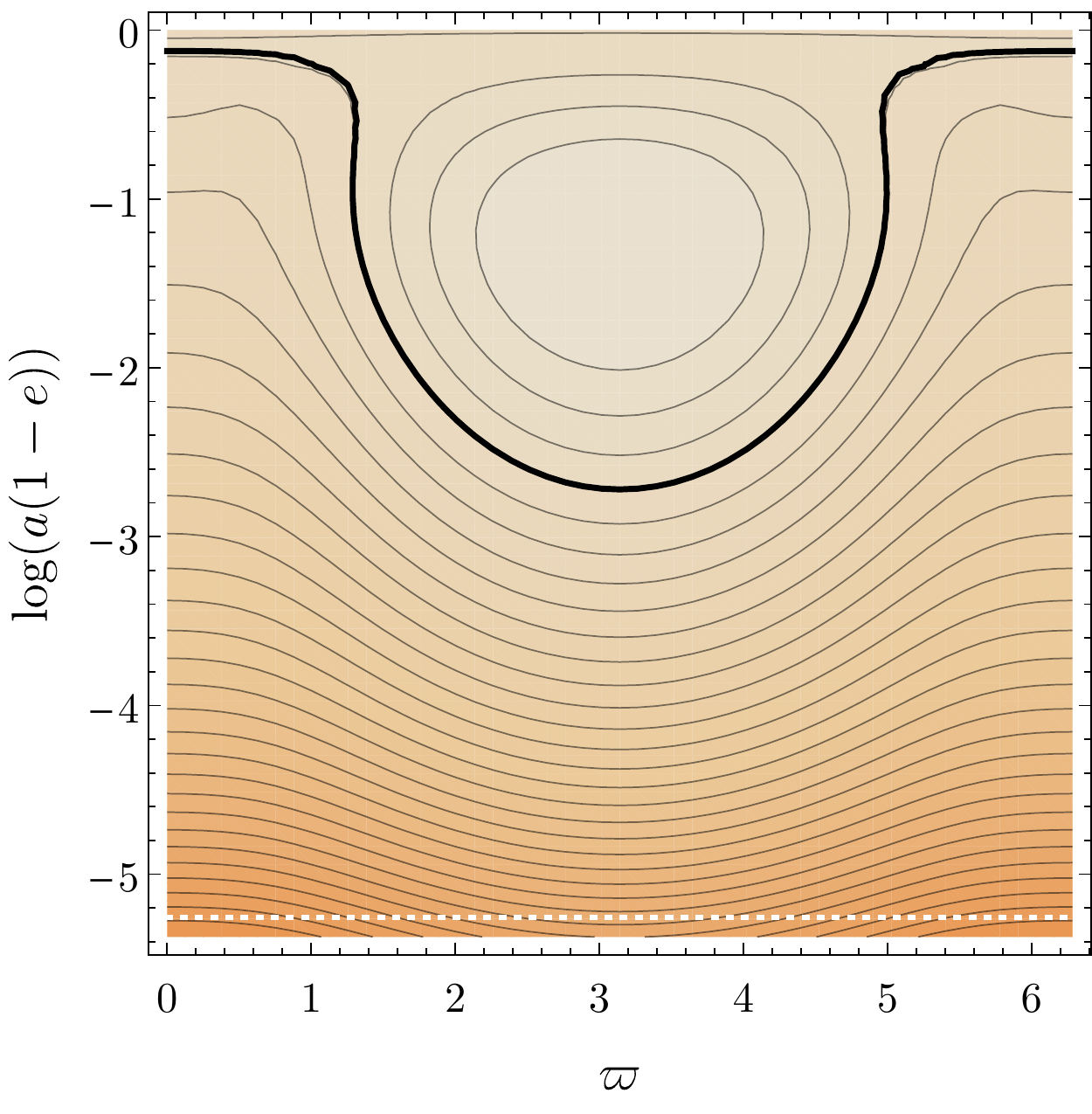}
\caption{Adding the General Relativity contribution.}
\end{subfigure}

\caption{Level curves of the Hamiltonians $\bar{\bar\Ha}$ (left panel) and $\bar{\bar\Ha}+\Ha_{GR}$ (right panel)
on the $(\VARPI, \log a_{per})$ plane for a test mass at $a_{res}=1~ \text{AU}$ in 4:1 mean motion resonance 
with an outer perturber with $\mu=3\times10^{-6}$, $\VARPI'=0$ and $e'=0.1$.
The mass of the parent star is set at $M_*= 1 M_\odot$.
The black solid line experiencing a significant change in $a_{peri}$ indicates the trajectory with 
the initial conditions $\VARPI=0$ and $e=0.05$.
The lower edge of the plot is at the location of the radius of the star, here taken to be the radius of the Sun ($R_\odot$).
The white dotted line indicates the location of the Roche limit, calculated using a density of the test particle of 
$\rho_{tp}=2~ \text{g}/\text{cm}^3$.
Note how the addition of the General Relativity potential reduces drastically the efficiency of the planetary perturbation 
in driving the test particle to collide with the star.
}
\label{fig:Figure9} 
\end{figure*}

\begin{figure*}
\centering
\begin{subfigure}[b]{0.45 \textwidth}
\centering
\includegraphics[scale=0.65
]{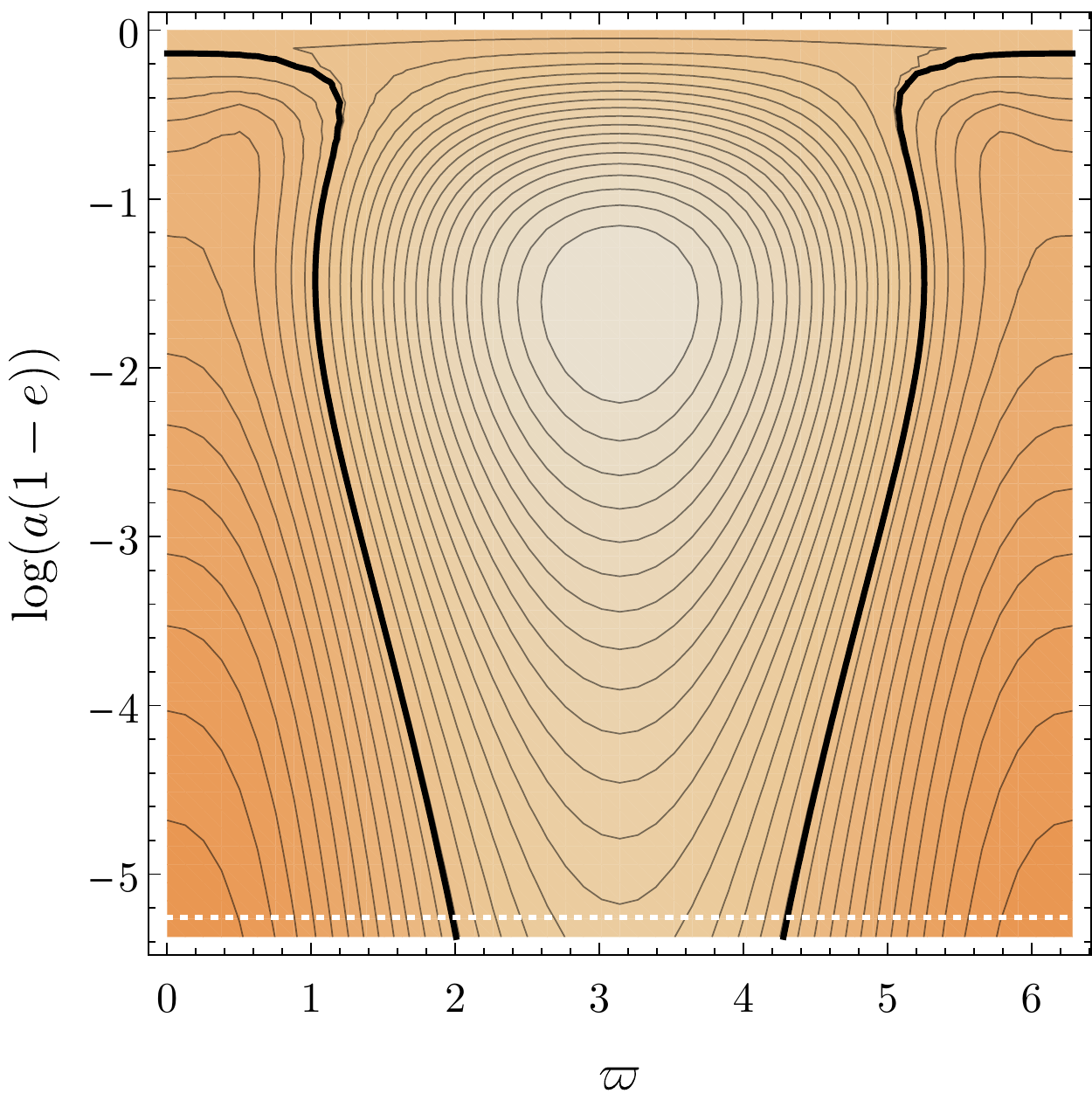}
\caption{Purely planetary case.}
\end{subfigure}
\begin{subfigure}[b]{0.45 \textwidth}
\centering
\includegraphics[scale=0.65
]{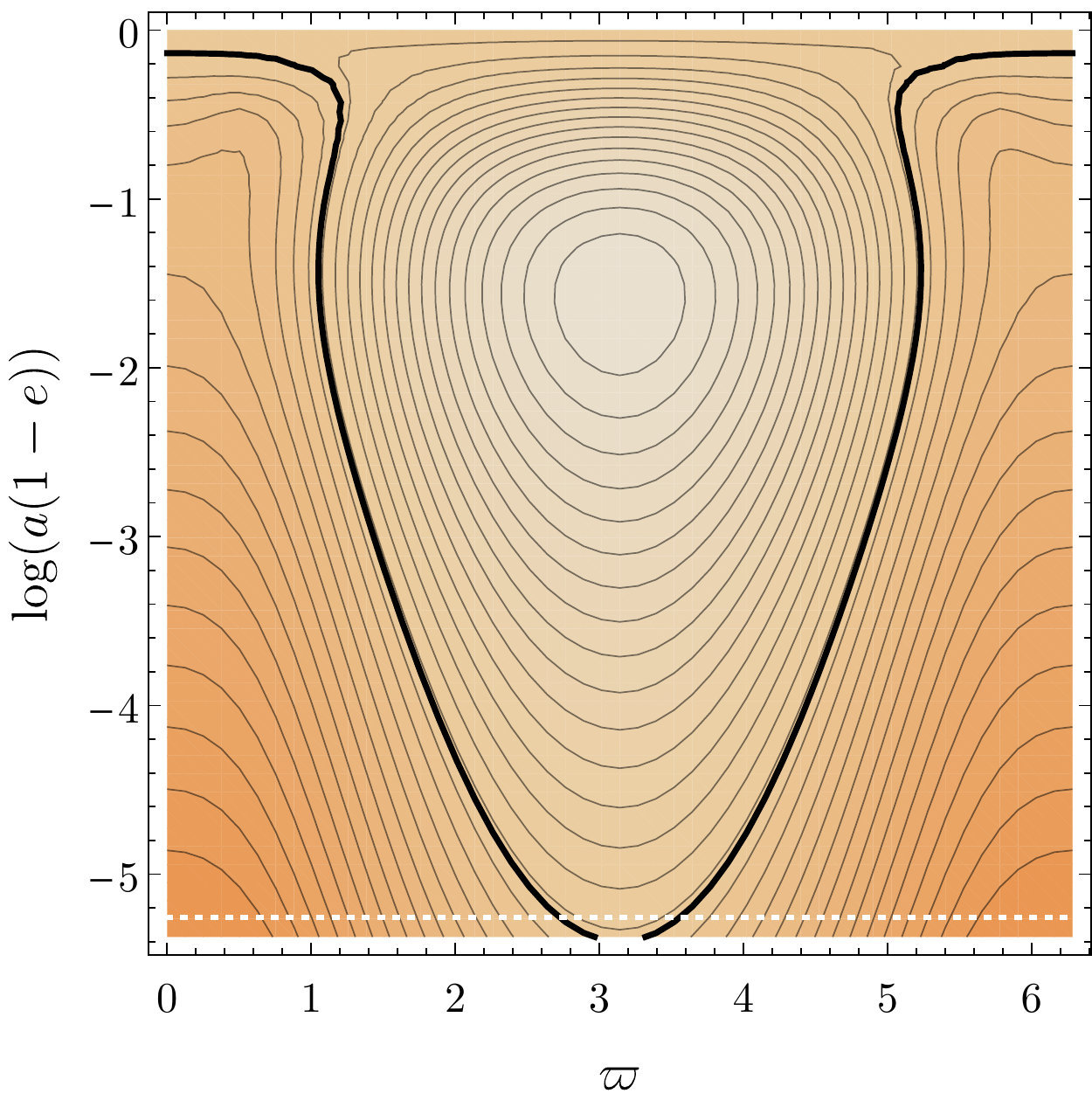}
\caption{Adding the General Relativity contribution.}
\end{subfigure}

\caption{Same as in Figure \ref{fig:Figure9}, except for $\mu=5\times10^{-5}$. 
In this case, a test mass starting at $\VARPI=0$ and $e=0.05$ can fall into the star even considering the General Relativity 
contribution.
Note that of course the level curves of the purely planetary Hamiltonian do not change with different values of $\mu$: 
as explained in the text, here $\mu$ only plays the role of setting the timescales of the evolution of the test particle, 
not the evolution itself.
}
\label{fig:Figure10} 
\end{figure*}

In Figure \ref{fig:Figure9} we show the level curves of the Hamiltonian $\bar{\bar\Ha}$ with $e'=0.1$, on the 
$(\VARPI,\log a_{peri})$ plane, both with and without the addition of the General Relativity contribution, 
for the case of $\mu=3\times 10^{-6}$. 
Here we choose the resonance location of the test particle $a_{res}$ at $1~ \text{AU}$.
We can clearly see that while in the purely planetary case the resonance is capable of pushing a test mass with a small
initial eccentricity $e\sim0.05$ onto a star-grazing orbit, this does not hold true when $\Ha_{GR}$ is introduced.
In Figure \ref{fig:Figure10} we repeat the calculation, this time with $\mu=5\times 10^{-5}$ and the same values for 
$a_{res}=1~ \text{AU}$ and $e'=0.1$, and we see that even with the General
Relativity contribution, test particles with initial small eccentricities are just about able to reach $a_{peri} \sim R_\odot$.
Because the thick curve in Figure \ref{fig:Figure10}(b) is almost tangent to the bottom of the figure at $\VARPI=\pi$, 
we deduce that the critical mass to achieve star-grazing orbits for this choice of $a'$ and $e'$ is close to $5\times 10^{-5}$ 
solar masses. 

The critical perturber mass $\mu_{crit}=m'_{crit}/M_*$ can be estimated as follows. 
For a test particle near a given mean-motion resonance (4:1) with an external perturber (of given $m', a', e'$), the
``secular'' planetary Hamiltonian can be written schematically as 
\begin{equation}
{\bar{\bar{\Ha}}}=-\Phi_0{\hat{\Ha}}(e,\VARPI), 
\end{equation} 
where 
\begin{equation}
\Phi_0\equiv {\GravC m'a^2\over {a'}^3}\propto {m'\over a}, 
\end{equation} 
and ${\hat{\Ha}}$ is dimensionless. 
Note that in the above equation, $a$ is really $a_0=a'/4^{2/3}$ (the value of $a$ in exact Keplerian resonance with the perturber). 
Suppose the test mass starts with an initial eccentricity $e_0\ll 1$ at $\varpi=0$. 
Its maximum eccentricity $e_{max}^{(0)}$ (achieved at $\VARPI=\pi$) is determined by 
\begin{equation}
{\hat{\Ha}}(e_0,0)-{\hat{\Ha}}(e_{max}^{(0)},\pi)=0.  
\end{equation}
The superscript ``$(0)$'' in $e_{max}^{(0)}$ indicates that this maximum eccentricity is obtained without any short-range 
force effect.

Now consider how ${\Ha}_{GR}$ affects $e_{max}$. We write
\begin{equation}
{\Ha}_{GR}=-{\Phi_{GR}\over \sqrt{1-e^2}},
\end{equation}
with 
\begin{equation}
\Phi_{GR}\equiv {3\GravC M_*\over a}{\GravC M_*\over a c^2}.
\end{equation}
Again, starting with an initial eccentricity $e_0\ll 1$ at $\varpi=0$, the maximum eccentricity $e_{max}$ of the test mass 
(achieved at $\varpi=\pi$) is estimated by 
\begin{equation}
\Phi_0 {\hat{\Ha}}(e_0,0)+\Phi_{GR} \simeq \Phi_0 {\hat{\Ha}}(e_{max},\pi)+{\Phi_{GR}\over \sqrt{1-e_{max}^2}}.
\label{eq:emax1}
\end{equation}
Assuming $1-e_{max}\ll 1$, equation (\ref{eq:emax1}) becomes
\begin{equation}
{|{\Ha}_{GR}|\over\Phi_0}={\Phi_{GR}\over\Phi_0}{1\over\sqrt{1-e_{max}^2}} \simeq
	 {\hat{\Ha}}(e_0,0)-{\hat{\Ha}}(e_{max},\pi) = : f.
\label{eq:emax2}
\end{equation}
This shows that $e_{max}$ depends on various parameters through the ratio $\Phi_{GR}/\Phi_0\propto M_*^2/(m'a)$.

Setting $a(1-e_{max})=R_{crit}$ in equation (\ref{eq:emax2}), we obtain the critical perturber
mass $m'_{crit}$ that allows the test particle to reach a certain pericenter distance $R_{crit}$:
\begin{equation}
\begin{split}
m'_{crit} &= \frac{3}{\sqrt{2}} \frac{1}{f} \frac{\GravC M_*^2}{c^2}\frac{1}{\sqrt{R_*}}a^{\!-1/2}
	\left(\frac{a(1-e_{max})}{R_*}\right)^{\!-1/2}\left(\frac{4}{1}\right)^{2} \\
	&\simeq 17 M_\oplus \left({f\over 0.1}\right)^{\!-1}
	\left(\!{M_*\over M_\odot}\!\right)^2
	\left({a\over{\rm AU}}\right)^{\!-1/2}\!
	\left({R_{crit}\over R_\odot}\right)^{\!-1/2}.
\end{split}
\end{equation}

Note that $f$ in general depends on $e_{max}$ and thus is a complicated function of $(R_{crit}/a)$.  
However, we can calculate its numerical value in the case depicted in Figures \ref{fig:Figure9} and \ref{fig:Figure10},
where we obtain $f \sim 0.1$.

\section{Conclusions}

In this paper, we have revisited the problem of resonant dynamics inside mean motion resonances in the restricted planar three-body 
problem, to determine to what extent planetary perturbations can effectively drive small bodies onto highly eccentric orbits 
and fall into the star or suffer tidal disruption.
While previous works employed series expansion of the Hamiltonian in powers of the eccentricities, or were limited by 
a first order development in $e'$ to the case of $e>e'$ and small $e'$
(where $e'$ and $e$ are the eccentricities of the planetary perturber and the test particle, respectively), 
we do not perform any expansions, thus making our results valid for a wider range of orbital configurations. 
We make use of the principle of adiabatic invariance to reduce the two-degree-of-freedom Hamiltonian \eqref{eq:AveragedHamiltonian} 
to the integrable Hamiltonian $\bar{\bar\Ha}$, which we study in the limit of vanishing amplitude of libration 
of $k'\LAMBDA'-k\LAMBDA$ in the $k':k= 2:1$, $3:1$ and $4:1$ mean motion resonances. 
We confirm the results of \cite{1996Icar..120..358B}, and show that for small $e'$ ($\lesssim 0.1$) 
the 2:1 and 3:1 resonances are not able to push test particles in initially nearly circular orbits 
onto star-grazing trajectories (Figures \ref{fig:Figure3}, \ref{fig:Figure4}), 
while the 4:1 resonance is extremely effective (Figure \ref{fig:Figure5}).
Moreover, we find that a larger value of $e'$ (=0.2-0.3) does not change this picture for the 2:1 and 3:1 resonances 
(Figures \ref{fig:Figure6}, \ref{fig:Figure7}), but makes the 4:1 resonance less effective, 
by generating a larger stable region of circulation around $e\sim0$ (Figure \ref{fig:Figure8}).
Finally, in the cases where the resonance is strong enough to generate star-grazing objects, we include the General Relativity
contribution to the Hamiltonian, which causes a fast precession of the pericenter while keeping the eccentricity constant, 
thereby suppressing the effectiveness of the planet's perturbation to generate extreme eccentricities (Figure \ref{fig:Figure9}). 
We note that, while the planetary mass only sets the timescales of the secular eccentricity evolution when the General Relativity 
effect is neglected, it now plays an important dynamical role, as it regulates the relative contribution of the 
purely Newtonian evolution and the impact of the post-Newtonian term. 
We then obtain, for a specific choice of semi-major axis and eccentricity of the perturber, an estimate on the minimum planetary 
mass needed to drive eccentricity growth of the test particle from $\sim0$ to $\sim1$.
An approximate analytic expression for this critical mass is also obtained.
In addition, we make available a Mathematica notebook which implements the calculations outlined in the paper, 
to allow the interested reader to examine the effect of secular dynamics inside mean motion resonances for other applications.

\section{Acknowledgments}
GP wishes to thank Cl\'ement Robert for the valuable  advice in improving the readability of Mathematica 
code. 
DL acknowledges support by the NASA grants NNX14AG94G and NNX14AP31G, and by a Simons Fellowship in Theoretical Physics from the Simons Foundation.
He also thanks the Laboratoire Lagrange OCA for hospitality.

\newcommand{\amp}{\&}


\begin{thebibliography}{9}

\bibitem[Beaug{\'e} et al.(2006)]{2006MNRAS.365.1160B} Beaug{\'e}, C., Michtchenko, T.~A., \& Ferraz-Mello, S.\ 2006, \mnras, 365, 1160 

\bibitem[Beust et al.(1991)]{1991A&A...241..488B} Beust, H., Vidal-Madjar, A., Ferlet, R., Lagrange-Henri, A.~M.\ 1991.\ The Beta Pictoris circumstellar disk. XI - New CA II absorption features reproduced numerically.\ Astronomy and Astrophysics 241, 488-492. 

\bibitem[Beust et al.(1990)]{1990A&A...236..202B} Beust, H., Vidal-Madjar, A., Ferlet, R., Lagrange-Henri, A.~M.\ 1990.\ The Beta Pictoris circumstellar disk. X - Numerical simulations of infalling evaporating bodies.\ Astronomy and Astrophysics 236, 202-216. 

\bibitem[Beust et al.(1989)]{1989A&A...223..304B} Beust, H., Lagrange-Henri, A.~M., Vidal-Madjar, A., Ferlet, R.\ 1989.\ The Beta Pictoris circumstellar disk. IX - Theoretical results on the infall velocities of CA II, AL III, and MG II.\ Astronomy and Astrophysics 223, 304-312. 

\bibitem[Beust and Morbidelli(1996)]{1996Icar..120..358B} Beust, H., Morbidelli, A.\ 1996.\ Mean-Motion Resonances as a Source for Infalling Comets toward {$\beta$} Pictoris.\ Icarus 120, 358-370. 

\bibitem[Bonsor et al.(2011)]{2011MNRAS.414..930B} Bonsor, A., Mustill, A.~J., \& Wyatt, M.~C.\ 2011, \mnras, 414, 930 

\bibitem[Debes et al.(2012)]{2012ApJ...747..148D} Debes, J.~H., Walsh, K.~J., \& Stark, C.\ 2012, \apj, 747, 148 

\bibitem[Debes and Sigurdsson(2002)]{2002ApJ...572..556D} Debes, J.~H., Sigurdsson, S.\ 2002.\ Are There Unstable Planetary Systems around White Dwarfs?.\ The Astrophysical Journal 572, 556-565. 

\bibitem[Marsden(1967)]{1967AJ.....72.1170M} Marsden, B.~G.\ 1967.\ The sungrazing comet group.\ The Astronomical Journal 72, 1170. 

\bibitem[Farihi(2016)]{2016NewAR..71....9F} Farihi, J.\ 2016.\ Circumstellar debris and pollution at white dwarf stars.\ New Astronomy Reviews 71, 9-34. 

\bibitem[Farinella et al.(1994)]{1994Natur.371..315F} Farinella, P., Froeschle, C., Froeschle, C., Gonczi, R., Hahn, G., Morbidelli, A., Valsecchi, G.~B.\ 1994.\ Asteroids falling onto the Sun.\ Nature 371, 315-317. 

\bibitem[Ferlet et al.(1987)]{1987A&A...185..267F} Ferlet, R., Vidal-Madjar, A., \& Hobbs, L.~M.\ 1987, The Beta Pictoris circumstellar disk. V - Time variations of the CA II-K line \aap, 185, 267 

\bibitem[Fontaine and Michaud(1979)]{1979ApJ...231..826F} Fontaine, G., Michaud, G.\ 1979.\ Diffusion time scales in white dwarfs.\ The Astrophysical Journal 231, 826-840. 

\bibitem[Gladman et al.(1997)]{1997Sci...277..197G} Gladman, B.~J., Migliorini, F., Morbidelli, A., Zappala, V., Michel, P., Cellino, A., Froeschle, C., Levison, H.~F., Bailey, M., Duncan, M.\ 1997.\ Dynamical lifetimes of objects injected into asteroid belt resonances.\ Science 277, 197-201. 

\bibitem[Greaves et al.(2016)]{2016MNRAS.461.3910G} Greaves, J.~S., Holland, W.~S., Matthews, B.~C., et al.\ 2016, \mnras, 461, 3910 

\bibitem[Jura(2003)]{2003ApJ...584L..91J} Jura, M.\ 2003.\ A Tidally Disrupted Asteroid around the White Dwarf G29-38.\ The Astrophysical Journal 584, L91-L94. 

\bibitem[Henrard(1993)]{Henrard1993} Henrard, J.\ 1993, The Adiabatic Invariant in Classical Mechanics. Dynamics Reported, 117-235, Springer.

\bibitem[Henrard \& Lamaitre(1983)]{1983CeMec..30..197H} Henrard J. and Lemaitre A.. 1983. A second fundamental model for resonance. Celest. Mech., 30, 197.

\bibitem[Henrard and Caranicolas(1990)]{1990CeMDA..47...99H} Henrard, J., Caranicolas, N.~D.\ 1990.\ Motion near the 3/1 resonance of the planar elliptic restricted three body problem.\ Celestial Mechanics and Dynamical Astronomy 47, 99-121. 

\bibitem[Koester et al.(2014)]{2014A&A...566A..34K} Koester, D., G{\"a}nsicke, B.~T., \& Farihi, J.\ 2014, \aap, 566, A34 

\bibitem[Krivov(1986)]{1986SvA....30..224K} Krivov, A.~V.\ 1986, \sovast, 30, 224 

\bibitem[Lagrange et al.(1987)]{1987A&A...173..289L} Lagrange, A.~M., Ferlet, R., Vidal-Madjar, A.\ 1987.\ The Beta Pictoris circumstellar disk. IV - Redshifted UV lines.\ Astronomy and Astrophysics 173, 289-292. 

\bibitem[Lemaitre(1984)]{1984CeMec..32..109L} Lemaitre, A.\ 1984.\ High-order resonances in the restricted three-body problem.\ Celestial Mechanics 32, 109-126. 

\bibitem[Liu et al.(2015)]{2015MNRAS.447..747L} Liu, B., Mu{\~n}oz, D.~J., \& Lai, D.\ 2015, \mnras, 447, 747 

\bibitem[Michtchenko et al.(2006)]{2006CeMDA..94..411M} Michtchenko, T.~A., Beaug{\'e}, C., \& Ferraz-Mello, S.\ 2006, Celestial Mechanics and Dynamical Astronomy, 94, 411 

\bibitem[Montgomery et al.(2015)]{2015AAS...22534919M} Montgomery, S.~L., Welsh, B., Bukoski, B., Strausbaugh, S.\ 2015.\ Exocomets and variable circumstellar gas absorption in the debris disks of nearby A-type stars.\ American Astronomical Society Meeting Abstracts 225, 349.19. 

\bibitem[Moons and Morbidelli(1993)]{1993CeMDA..57...99M} Moons, M., Morbidelli, A.\ 1993.\ The main mean motion commensurabilities in the planar circular and elliptic problem.\ Celestial Mechanics and Dynamical Astronomy 57, 99-108. 

\bibitem[Moons and Morbidelli(1995)]{1995Icar..114...33M} Moons, M., Morbidelli, A.\ 1995.\ Secular resonances inside mean-motion commensurabilities: the 4/1, 3/1, 5/2 and 7/3 cases..\ Icarus 114, 33-50. 

\bibitem[Morbidelli(2002)]{2002mcma.book.....M} Morbidelli, A.\ 2002, Modern celestial mechanics : aspects of solar system dynamics, by Alessandro Morbidelli.~London: Taylor {\amp} Francis, 2002, ISBN 0415279399,  

\bibitem[Murray \& Dermott(2000)]{2000ssd..book.....M} Murray, C.~D., \& Dermott, S.~F.\ 2000, ``Solar System Dynamics'' Cambridge, UK: Cambridge University Press.

\bibitem[Petrovich \& Mu{\~n}oz(2017)]{2017ApJ...834..116P} Petrovich, C., \& Mu{\~n}oz, D.~J.\ 2017, \apj, 834, 116 

\bibitem[Sorelli et al.(1996)]{1996A&A...309..155S} Sorelli, C., Grinin, V.~P., Natta, A.\ 1996.\ Infall in Herbig Ae/Be stars: what NA D lines tell us..\ Astronomy and Astrophysics 309, 155-162. 

\bibitem[Vauclair and Fontaine(1979)]{1979ApJ...230..563V} Vauclair, G., Fontaine, G.\ 1979.\ Convective mixing in helium white dwarfs.\ The Astrophysical Journal 230, 563-569. 

\bibitem[Welsh \& Montgomery(2013)]{2013PASP..125..759W} Welsh, B.~Y., \& Montgomery, S.\ 2013, \pasp, 125, 759 

\bibitem[Wisdom(1985)]{1985Icar...63..272W} Wisdom, J.\ 1985.\ A perturbative treatment of motion near the 3/1 commensurability.\ Icarus 63, 272-289. 

\bibitem[Wisdom(1983)]{1983Icar...56...51W} Wisdom, J.\ 1983.\ Chaotic behavior and the origin of the 3/1 Kirkwood gap.\ Icarus 56, 51-74. 

\bibitem[Zuckerman et al.(2003)]{2003ApJ...596..477Z} Zuckerman, B., Koester, D., Reid, I.~N., H{\"u}nsch, M.\ 2003.\ Metal Lines in DA White Dwarfs.\ The Astrophysical Journal 596, 477-495. 


\end{thebibliography}
\end{document}